\documentclass[a4paper,fleqn]{cas-dc}

\usepackage[numbers]{natbib}
\usepackage{xurl}
\usepackage{tabularx}
\usepackage{amsthm}
\usepackage{amsmath}
\usepackage{amssymb}
\usepackage{enumitem}

\usepackage{listings}

\usepackage{tabularx}
\usepackage{multirow}
\usepackage{booktabs}

\usepackage{siunitx}
\usepackage{numprint}

\makeatletter
\def\@@@resptofloat#1#2{%
  \gdef\@fpsadd{~}%
  \@ifnextchar[{\@@resptofloat{#1}{#2}}{\@resptofloat{#1}{#2}[!htbp]}%
}
\makeatother

\def\reified#1{\mbox{``}#1\mbox{''}}

\def\Dimspace{\mathfrak{D}_\Omega}

\def\indiv#1{\mathsf{#1}}
\def\concept#1{\mathsf{#1}}
\def\role#1{\mathsf{#1}}
\def\isa{\sqsubseteq}
\def\CKB{\mathfrak{K}}

\def\I{\mathcal{I}}

\newcommand{\IC}{\mathfrak{I}}

\newcommand{\stru}[1]{\langle #1 \rangle}

\renewcommand{\dim}{\vdash}

\newcommand{\Ccal}{{\cal C}}

\newcommand{\Lcal}{{\cal L}}
\newcommand{\Scal}{{\cal S}}
\newcommand{\Mcal}{{\cal M}}

\newcommand{\Tcal}{{\cal T}}

\newcommand{\cov}[1]{\preceq}

\newcommand{\nop}[1]{}

\newtheorem{definition}{Definition}
\newtheorem{proposition}{Proposition}

\setitemize{label=--,leftmargin=*}
\setenumerate{label=(\roman*).,leftmargin=*}
\newcommand{\EndEx}{\mbox{}~\hfill$\Diamond$} 

\newcommand{\Gl}{\mathfrak{G}}

\newcommand{\ml}[1]{\mathsf{#1}} 

\newcommand{\cn}{{\sf cn}}
\newcommand{\Idx}{\text{\textit{Idx}}}	

\theoremstyle{definition}
\newtheorem{example}{Example}

\definecolor{iGrey}{RGB}{128,128,128}

\newcommand{\Doc}{Doc} 

\def\tsc#1{\csdef{#1}{\textsc{\lowercase{#1}}\xspace}}
\tsc{WGM}
\tsc{QE}
\tsc{EP}
\tsc{PMS}
\tsc{BEC}
\tsc{DE}

\ExplSyntaxOn
\cs_gset:Npn \__first_footerline: {
  \group_begin: \small \sffamily \__short_authors: \group_end:
}
\ExplSyntaxOff

\begin{document}
\let\WriteBookmarks\relax
\shorttitle{Knowledge Graph Lakehouse: Managing Enterprise Knowledge Graphs at Scale}
\shortauthors{B. Ahmad et~al.}

\title[mode=title]{Knowledge Graph Lakehouse: A Cloud-Native Architecture for Managing Enterprise Knowledge Graphs at Scale}

\author[1]{B. Ahmad}[bioid=1,
                     orcid=0000-0001-9610-1516]
\cormark[1]
\ead{ahmad@dke.uni-linz.ac.at}
\credit{Conceptualization, Methodology, Formal analysis, Software, Investigation, Writing -- original draft, Visualization}

\affiliation[1]{organization={Institute of Business Informatics -- Data \& Knowledge Engineering, Johannes Kepler University Linz},
                addressline={Altenberger Str. 69}, 
                postcode={4040}, 
                postcodesep={}, 
                city={Linz},
                country={Austria}}

\author[1]{C. G. Schuetz}[bioid=2,
                          orcid=0000-0002-0955-8647]
\credit{Conceptualization, Methodology, Supervision, Writing -- review \& editing}

\author[2]{L. Bozzato}[bioid=3,
                       orcid=0000-0003-1757-9859]
\credit{Formal analysis, Methodology, Writing -- review \& editing}

\affiliation[2]{organization={Dipartimento di Scienze Teoriche e Applicate, Università degli Studi dell'Insubria},
                addressline={Via O. Rossi 9}, 
                postcode={21100}, 
                postcodesep={}, 
                city={Varese},
                country={Italy}}

\cortext[cor1]{Corresponding author}


\begin{abstract}
Modern data architectures for integrating heterogeneous data sources to discover actionable knowledge include the data warehouse, the data lake, and the data lakehouse, which combines the flexibility of the data lake with the structured data management capabilities of the data warehouse through metadata, indexes, and caching.
Enterprise knowledge graphs (KGs) offer an alternative approach to such integration for the purposes of data analysis, but handling them at scale must address the Three Vs of big data: volume, velocity, and variety.
Contextualized KGs (CKGs) meet part of that challenge by enriching facts with contextual information, yet many applications require only a context-specific subset of that knowledge.
We propose the Knowledge Graph Lakehouse (KG Lakehouse), a cloud-native microservice architecture that adapts the data lakehouse paradigm for management of CKGs at scale.
The KG Lakehouse indexes incoming source documents along configurable context dimensions and stores these documents in a data lake, from which application-specific CKGs are constructed on demand instead of maintaining a single, monolithic KG.
We realize the proposed architecture as a proof of concept, with independently deployable services built from established cloud technology; context dimensions and format-specific document handling enter as configuration, leaving the architecture independent of the application domain.
We evaluate the implemented system in experiments on a distributed cluster deployment.
\end{abstract}


\maketitle

\section{Introduction}\label{sec:intro}

Modern data architectures for integrating heterogeneous data sources to discover actionable knowledge include the \emph{data warehouse}, the \emph{data lake}, and the \emph{data lakehouse} (see~\cite{Harby2022} for an overview).
A data warehouse stores structured data that is clean and consistent~\cite[p.~63]{vaisman2022}, optimized for recurring analytical tasks, with analytical results typically summarized in reports or dashboards~\cite[p.~625]{vaisman2022}.
A data lake retains raw or minimally processed data to support flexible exploratory analysis and future analytical use cases~\cite[p.~624]{vaisman2022}.
Although flexible, the data lake's \emph{schema-on-read} approach requires complex cleaning and transformation before analysis~\cite[p.~625]{vaisman2022}, increasing processing overhead and complicating data management for large datasets.
The data lakehouse~\cite{Armbrust2021} combines the flexibility of the data lake with the structured data management capabilities of the data warehouse by leveraging metadata, indexes, and caching to support complex data analysis while providing storage efficiency and query flexibility.

Enterprise knowledge graphs have been proposed as an alternative approach to integrating heterogeneous data sources; see~\cite{Sequeda2022} for one such proposal.
A \emph{knowledge graph} (KG) comprises facts about real-world entities and their relationships. 
KGs have been described as ``large networks of entities, their semantic types, properties, and relationships''~\cite{krotzsch2016}, ``a set of interconnected typed entities and their attributes``~\cite[p.~1]{gomez-perez2017} with possibly arbitrary relationships~\cite{paulheim2017}. 
Apart from the instance-level facts, or assertional knowledge (ABox)~\cite{paulheim2017}, a KG also often comprises terminological or ontological knowledge (TBox) representing ``the vocabulary used in the knowledge graph''~\cite[p.~2]{gomez-perez2017} and enabling ontological reasoning over the facts~\cite{bellomarini2017}. 
Unlike open KGs, an \emph{enterprise KG} is ``typically internal to a company and applied for commercial use-cases''~\cite[p.~71:4]{Hogan2021}, containing relevant knowledge for an organization~\cite{bellomarini2017}.
Due to their versatility, enterprise KGs ``are a natural data model for data integration''~\cite[p.~10]{Sequeda2022}.
By combining assertional and ontological knowledge, enterprise KGs capture both data and their semantics, making it easier for data consumers to discover actionable knowledge~\cite{Sequeda2022}.

Effective deployment of enterprise KGs relies on a \emph{KG management system} (KGMS), which not only provides ontological reasoning capabilities but also supports large-scale data management and analytical tasks~\cite{bellomarini2017}.
The knowledge represented in enterprise KGs expands and evolves over time, incorporating information from various structured and unstructured sources and existing at multiple levels of granularity and abstraction~\cite{noy2019}.
Data sources continuously provide large volumes of heterogeneous information, often at high velocity, that must be integrated into the enterprise KG.
The KGMS must therefore be able to handle the \emph{Three~Vs} of big data~\cite{Laney2001}: \emph{volume}, \emph{velocity}, and \emph{variety}.

At the conceptual level, \emph{contextualized KGs} \cite{Serafini2012,BozzatoES:18} provide a model for managing heterogeneous and evolving knowledge by enriching facts with contextual information.
In a contextualized KG (CKG), facts are typically organized into subgraphs, each corresponding to a distinct context.
To represent CKGs, the \emph{Knowledge Graph OLAP} (KG-OLAP) model~\cite{schuetz2021} employs a multidimensional context model with hierarchically organized dimensions.
The dimensions capture different contextual characteristics, e.g., geographic scope, temporal applicability, and topic.
The hierarchical organization of the dimensions supports the representation of contexts at different levels of granularity (e.g., sector and country for geographic scope, and day and month for temporal applicability).
Dedicated operations, adapted to KGs from traditional \emph{online analytical processing} (OLAP), enable the navigation and combination of contexts as well as the transformation of the graph structure as required for specific analytical tasks.

\begin{figure*}
     \includegraphics[width=\textwidth]{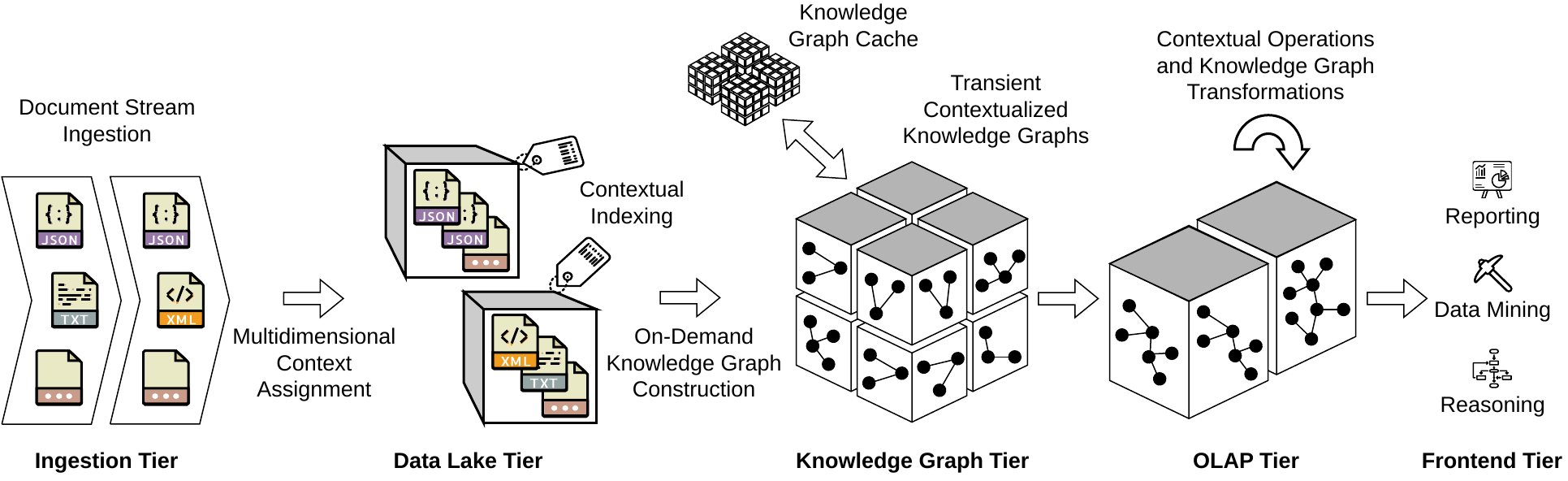}
    \caption{Conceptual overview of the KG Lakehouse}
    \label{fig:conceptual_architecture}
\end{figure*}

At the architectural level, a cloud-native microservice architecture provides the modularity, elasticity, and scalability required to manage CKGs at scale.
Different services can be independently developed, deployed, and horizontally scaled according to workload requirements.
New services can be introduced to integrate additional data sources without affecting existing components.
A microservice architecture may further leverage the contextual partitioning of knowledge in CKGs to enable parallel processing of independent contexts for improved performance.

Even with CKGs and a cloud-native microservice architecture, maintaining a single, monolithic enterprise KG at scale remains challenging and introduces overhead that many applications do not require.
A KG typically covers a variety of topics rather than focusing exclusively on a single aspect of the real world~\cite{paulheim2017}.
For specific tasks, however, only a subset of the knowledge represented in an enterprise KG may be relevant, since ``most people do not care about all entities that exist in the world, but rather a small fraction or subset that is personally relevant to them''~\cite[p.~43]{noy2019}.
For example, in air traffic management, which serves as a motivating use case and setting of the running example throughout this paper, a KG may represent knowledge about the air traffic network, including infrastructure, airspace, operational events, weather conditions, and flight plans~\cite{Schuetz2018}.
A pilot preparing for a flight from Munich to Vienna only requires knowledge of weather conditions and operational events, such as airspace closures and operational restrictions of navigation aids, that are relevant to the airspace sectors along the planned route on the day of the flight, rather than knowledge about the entire European air traffic network~\cite{Schuetz2019}.
An air traffic controller, on the other hand, requires the same contextual knowledge, together with the flight plans as well as positions of aircraft operating in the supervised airspace sector and its adjacent sectors during a specific period of responsibility.
Hence, maintaining a single, monolithic KG is both impractical and unnecessary for many applications, which instead require only context-specific subsets of the available knowledge.

We propose the \emph{Knowledge Graph Lakehouse} (KG Lakehouse), a cloud-native architecture that adapts the data lakehouse paradigm~\cite{Armbrust2021} for CKG management.
Figure~\ref{fig:conceptual_architecture} provides a conceptual overview of the proposed architecture.
Documents of various types continuously arrive from different sources in a data stream.
During ingestion, the documents are analyzed and indexed along configurable context dimensions (e.g., location, time, and topic) before being stored in the data lake.
Instead of maintaining a single, monolithic KG, the KG Lakehouse maintains a contextual index of source documents from which application-specific CKGs are dynamically constructed as required by the analytical tasks.
In addition to contextual indexing, the KG Lakehouse maintains a cache of previously built CKGs  to reduce the overhead of repeated construction.
Contextual operations, which allow for selecting and merging contexts, and transformation operations facilitate adaptation of CKGs to the requirements of specific frontend applications, including reporting, data mining, and rule-based reasoning.

The proposed architecture is independent of any particular application domain.
The context dimensions are configurable, the format-specific analysis of documents is modular.
We illustrate and motivate the proposed architecture on the case of air traffic management, which serves as the setting for our running example and evaluation.
We stress, though, that the implementation can also be configured for other application domains.

The main contributions of this paper are as follows.
\begin{itemize}
    \item \emph{Architecture.} We specify the KG Lakehouse, a cloud-native architecture for CKG management that exploits the context-specific, multi-perspective nature of knowledge to construct tailored CKGs on demand instead of maintaining a single, monolithic KG.
    We formalize the architecture in terms of the KG-OLAP model. 
    The KG-OLAP model defines a cube as a mapping from contexts to knowledge modules. 
    The KG Lakehouse defines that mapping over the indexed source documents, so that constructing a CKG on demand computes, for the requested contexts, the same modules that a KG-OLAP cube materialized in advance from the same documents would hold. 

    \item \emph{Implementation.} We realize the architecture as a proof of concept, implemented as a set of independently deployable services built from established cloud technologies and operated on a container platform.
    The implementation demonstrates feasibility of the architecture.
    The modular design makes the implementation independent of the application domain: the context dimensions and the format-specific handling of documents are supplied as configuration, so that serving another domain is a matter of declaration rather than of reworking the services.
    The source code and the deployment configuration are openly available~\cite{kglakehouse}.

    \item \emph{Evaluation.} We evaluate the implemented system in five experiments on a distributed cluster deployment, which test the system along the \emph{Three Vs} of big data.
    The experiments show that ingestion and indexing keep pace with arriving documents, that the constructed CKGs pass independent correctness checks at every level of the context hierarchies, and that the cost of constructing a CKG follows the scope of the CKG rather than the size of the corpus.
    A further experiment compares on-demand construction with a cube that is materialized in advance, in run time and in memory consumption, for the same requested knowledge.
    The experimental setup, the scripts, and the measured data are openly available~\cite{kglakehouseeval}.
\end{itemize}

Related work focuses on different aspects along the big data analysis pipeline, including graph storage layouts over data lakes, building graphs on streaming data, or distributed and parallel query processing, which serve an already existing graph.
Federated and virtual knowledge graphs avoid maintenance of a permanent copy but answer each query by delegating the query to remote endpoints or to  (relational) sources linked by a mapping.
The KG Lakehouse, however, records the contexts that the content of each raw document is valid in or relevant for.
Once a request for a KG specifies the required contexts, the KG Lakehouse retrieves the documents relevant for the contexts and lifts knowledge from those documents.

The remainder of this paper is organized as follows.
Section~\ref{sec:case} presents a use case from Air Traffic Management serving as a running example.
Section~\ref{sec:kgolap} revisits the multidimensional context model of KG-OLAP for the representation of CKGs.
Section~\ref{sec:architecture} describes and formalizes the proposed architecture.
Section~\ref{sec:poc} presents a proof-of-concept implementation.
Section~\ref{sec:evaluation} reports the results of the experimental evaluation.
Section~\ref{sec:related} reviews related work.
Section~\ref{sec:conc} concludes the paper.

\section{Use Case: Air Traffic Management}\label{sec:case}

We demonstrate the KG Lakehouse using the case of air traffic management (ATM), where the knowledge required by different stakeholders for different tasks is inherently context-specific~\cite{Schuetz2018}, must be delivered under time-critical conditions, and relies on information sources exhibiting the Three Vs of big data.
To ensure safe and efficient air traffic operations, various stakeholders, including pilots and air traffic controllers, require the right information for the right context at the right time.
This information originates from a large number of highly heterogeneous sources describing different aspects of the air traffic network, producing large volumes of data that are continuously updated.

Different information service providers distribute ATM information through the System Wide Information Management (SWIM) platform~\cite{swim}.
Permanent and temporary updates on infrastructure and airspace are published by air navigation service providers (ANSPs), superseding baseline information and previously disseminated updates.
For example, a navigation aid may become temporarily unavailable, resulting in a change of its operational status for a specific period, or its assigned frequency may change permanently.
Construction work near an airport's runway may constitute a potential operational hazard.
Meteorological service providers publish regular weather observations and forecasts, as well as updates on adverse weather conditions.
For example, an aerodrome reports multiple weather observations per day, and thunderstorms are significant weather events of interest for air traffic operations.
Flight plans are submitted and updated throughout their life cycle, depending on weather conditions and traffic demand.
Other information services provide aircraft positions derived from broadcasts sent via Automatic Dependent Surveillance–Broadcast (ADS-B), resulting in a continuous stream of surveillance information.

ATM information is exchanged between services and applications following various standardized exchange models. 
The Aeronautical Information Exchange Model (AIXM)~\cite{AIXM2019} represents information about airport and navigation infrastructure, airspace, and operational events, which are commonly disseminated through Notices to Airmen (NOTAMs).
The ICAO Meteorological Information Exchange Model (IWXXM)~\cite{IWXXM2019} represents meteorological observations and forecasts, which are disseminated through Meteorological Aerodrome Reports (METARs), Terminal Aerodrome Forecasts (TAFs), and Significant Meteorological Information (SIGMET) messages.
The Flight Information Exchange Model (FIXM)~\cite{FIXM2019} represents flight plans, including planned flight trajectories.
In addition, aircraft continuously broadcast their positions using Automatic Dependent Surveillance–Broadcast (ADS-B), providing information about their actual trajectories.
These messages are commonly encoded as XML or JSON documents.

KGs may serve as an integration layer for the diverse ATM information exchanged over SWIM. 
An ATM KG may comprise the knowledge about the world as relevant for air traffic operations.
For example, in the AISA~\cite{aisa} and AWARE~\cite{aware} projects, which investigate the integration of AI capabilities into the controllers' ATM software, a KG serves as an integrated data layer.
The information required for various tasks, e.g., supervising a particular airspace sector or conducing a particular flight, and subtasks, e.g., predicting trajectories of individual aircraft or predicting flight delay, can be extracted from the ATM KG.

In the European air traffic network, tens of thousands of flights are conducted every day~\cite{traffic}, connecting thousands of aerodromes of different sizes and traversing a large number of airspace sectors.
A single, monolithic ATM KG representing the operational environment relevant for air traffic operations across Europe may therefore grow to a substantial size.
The continuous arrival of updates from heterogeneous sources makes maintaining the ATM KG up to date challenging, particularly given the time-critical nature of ATM information, where stakeholders require relevant information immediately.

Many ATM applications do not require the entire ATM KG simultaneously, making the maintenance of a single, monolithic ATM KG unnecessary for many use cases.
For example, a pilot briefing for a specific flight only requires the subset of knowledge about infrastructure, airspace, and weather relevant to that flight~\cite{Steiner2016}.
Similarly, the training and application of machine learning models typically require only task-specific subsets of knowledge.
For example, flight-delay prediction models are often trained for individual aerodromes~\cite{bardach2020}.

\section{Multidimensional Context Model}
\label{sec:kgolap}

For the representation of CKGs, we adopt the KG-OLAP framework~\cite{schuetz2021}. 
The KG-OLAP framework organizes knowledge within a multidimensional space of contexts\allowbreak---a contextual (hyper-)cube comprising multiple cells, with each cell representing a context. 
The KG-OLAP model, which is based on description logic, distinguishes two layers: the \emph{upper layer}, which defines the multidimensional schema of the cube and the context-independent knowledge, and the \emph{lower layer}, which contains the knowledge associated with the individual cells.

In this section, we revisit the multidimensional model of the KG-OLAP framework~\cite{schuetz2021}, which serves as the basis for the formalization of our notion of CKGs and their derivation from ingested documents. 
For reference, Table~\ref{tab:notation} gives an overview of the notation elements used in the formalization throughout this work.

\begin{table}
\caption{Overview of the notation used in the formalization}
\label{tab:notation}
\centering
\small
\begin{tabularx}{\columnwidth}{lX} 
\toprule
\textbf{Symbol} & \textbf{Meaning} \\
\midrule
$\Omega, \Sigma$ & Cube and object vocabularies \\
$\Lcal_\Sigma$ & Object language over $\Sigma$ \\
$\Tcal_\Sigma$ & Global TBox in $\Lcal_\Sigma$ \\
$D, F$ & Sets of dimensions and cell names \\
$D_A, L_A$ & Dimension members and levels for dimension $A$, with $lev: D_A \to L_A$ \\
$\prec_A, \prec^L_A$ & Member and level ordering relations \\
$\Dimspace$ & Multidimensional space of dimensional vectors \\
$\mathbf{d}$ & Dimensional vector (coordinates) in $\Dimspace$ \\
$\cn(\mathbf{d})$ & Bijective mapping from coordinates to cells $F$ \\
$\preceq$ & Coverage relation between vectors/cells \\
$\Ccal$ & cells-to-module mapping \\
$\Gl$ & Upper layer: coverage, cell-to-module mapping $\Ccal$, and global TBox $\Tcal_\Sigma$ \\
$K_c$ & Local knowledge module (ABox) of cell $c$ \\
$\CKB$ & KG-OLAP cube $\stru{\Gl, K_F}$ \\
$\Doc$ & Set of ingested documents \\
$\Idx$ & Multidimensional index, $\Idx \subseteq F \times \Doc$ \\
$\phi$ & Extraction function: document to raw data items \\
$\kappa$ & Context assignment: raw data item to a set of dimensional vectors, $\kappa: \mathcal{R}_{\mathcal{S}} \to 2^{\Dimspace}$ \\
$\mathcal{R}_{\mathcal{S}}$ & Set of raw data items \\
$\lambda_{S,\Sigma}$ & Semantic lifting: raw data item to assertions \\
$RD_c$ & Raw data items associated with cell $c$ \\
$\mathbf{e}, \mathbf{l}$ & Scope vector and level assignment of a KG specification \\
$\rho_{\mathbf{l}}$ & Coordinate coarsening to the levels of $\mathbf{l}$ \\
$M_c$ & TBox-extended module, $M_c = K_c \cup \Tcal_\Sigma$ \\
\bottomrule
\end{tabularx}
\end{table}

A cube vocabulary $\Omega = (\mathrm{NR}_\Omega, \mathrm{NC}_\Omega, \mathrm{NI}_\Omega)$ consists of atomic roles $\mathrm{NR}_\Omega$, concepts $\mathrm{NC}_\Omega$, and individuals $\mathrm{NI}_\Omega$. 
The cube vocabulary also defines a set of dimensions $D \subseteq \mathrm{NR}_\Omega$ and levels $L \subseteq \mathrm{NI}_\Omega$. For each dimension $A \in D$, the members $D_A \subseteq \mathrm{NI}_\Omega$ are partially ordered by $\prec_A$ in a tree hierarchy with a unique root $all_A$. 
A mapping $lev: D_A \to L_A$ assigns members to levels, which are themselves ordered by $\prec^L_A$.
In the following we will also use $\preceq_A$ and $\preceq^L_A$ to denote the non-strict variant of
the $\prec_A$ and $\prec^L_A$ relations.

Cells of the cube are identified by their
coordinates in the multidimensional space: 
the coordinates of a cell for all the 
dimensions are represented as a 
dimensional vector. 
Formally, with $|D| = k$, we define a \emph{dimensional vector} as the set
\begin{center}
  $\mathbf{d} = \{ A_1 := d_1, \dots, A_k := d_k \}$
\end{center}
where every $d_j \in D_{A_j}$, with $1 \leq j \leq k$.
The value given by $\mathbf{d}$ to the dimension $A$ is denoted with $d_A$.

\begin{definition}[Multidimensional space and cells]
The \emph{multidimensional space} $\Dimspace$ is the set of all dimensional vectors $\mathbf{d} = \{A_1 := d_1, \dots, A_k := d_k\}$ for dimensions $A_1 \dots A_k \in D$. The set of \emph{cell names} $F \subseteq \mathrm{NI}_\Omega$ provides unique identifiers for these coordinates via a bijective function\linebreak
$\cn: \Dimspace \to F$; we denote by $\cn^-$ the inverse of $\cn$.
\end{definition}

\begin{figure}
    \centering
    \includegraphics[width=0.48\textwidth]{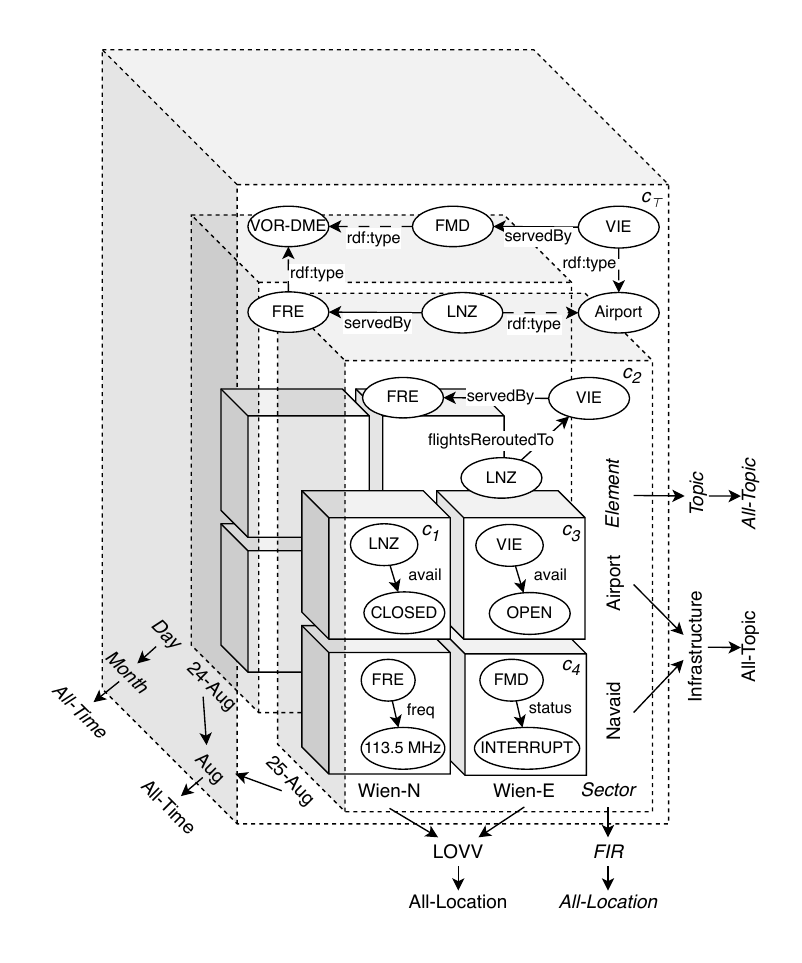}
    \caption{KG-OLAP cube for the ATM running example}
    \label{fig:cube-example}
\end{figure}

\begin{example}[Multidimensional space and cells]
\label{ex:cells}
Figure~\ref{fig:cube-example} shows a KG-OLAP cube for the ATM use case of Section~\ref{sec:case} with dimensions $D = \{\ml{Time}, \ml{Location}, \ml{Topic}\}$.
The example simplifies the ATM cube schema of the implementation (Section~\ref{sec:poc}) to three-level hierarchies with generic level and member names.
In the $\ml{Location}$ dimension, the members $\ml{Wien\text{-}N}$ and $\ml{Wien\text{-}E}$ at level $\ml{Sector}$ roll up to $\ml{LOVV}$, the Austrian flight information region, at level $\ml{FIR}$, and further to the root member $\ml{All\text{-}Location}$: $\ml{Wien\text{-}N} \prec_{\ml{Location}} \ml{LOVV} \prec_{\ml{Location}} \ml{All\text{-}Location}$.
Likewise, the days $\ml{24\text{-}Aug}$ and $\ml{25\text{-}Aug}$ at level $\ml{Day}$ roll up to $\ml{Aug}$ at level $\ml{Month}$ and then to $\ml{All\text{-}Time}$, and the topics $\ml{Airport}$ and $\ml{Navaid}$ at level $\ml{Element}$ roll up to $\ml{Infrastructure}$ and then to $\ml{All\text{-}Topic}$.
The dimensional vector $\mathbf{d}_1 = \{\ml{Time} := \ml{25\text{-}Aug},\, \ml{Location} := \ml{Wien\text{-}N},\, \ml{Topic} := \ml{Airport}\}$ identifies the cell $c_1 = \cn(\mathbf{d}_1)$, shown at the front left of the figure.
The vector $\mathbf{d}_2 = \{\ml{Time} := \ml{25\text{-}Aug},\, \ml{Location} := \ml{LOVV},\, \ml{Topic} := \ml{Infrastructure}\}$ identifies the covering cell $c_2 = \cn(\mathbf{d}_2)$, and the vector of root members $\{\ml{Time} := \ml{All\text{-}Time},\, \ml{Location} := \ml{All\text{-}Location},\, \ml{Topic} := \ml{All\text{-}Topic}\}$ identifies the most general cell $c_\top$.\EndEx
\end{example}

The coverage relation $\preceq$ represents a hierarchical organization of cells: the relation on cells is induced by the member orderings $\preceq_A$.
\begin{definition}[Context coverage]
For two vectors $\mathbf{d}, \mathbf{e} \in \Dimspace$, $\mathbf{d} \preceq \mathbf{e}$ holds if $d_A \preceq_A e_A$ for every dimension $A \in D$. 
For the corresponding cells $c_1 = \cn(\mathbf{d})$ and $c_2 = \cn(\mathbf{e})$, we write $c_1 \preceq c_2$ to indicate that $c_1$ is more specific than $c_2$ (and we say that $c_2$ \emph{covers} $c_1$).
\end{definition}
\noindent
As for the member orderings, we write $\mathbf{d} \prec \mathbf{e}$, and correspondingly $c_1 \prec c_2$, for the strict variant of the relation (i.e., if the two compared elements are distinct).

\begin{example}[Context coverage]
\label{ex:coverage}
Continuing Example~\ref{ex:cells},\linebreak $\mathbf{d}_1 \preceq \mathbf{d}_2$ holds because $\ml{25\text{-}Aug} \preceq_{\ml{Time}} \ml{25\text{-}Aug}$, $\ml{Wien\text{-}N} \preceq_{\ml{Location}} \ml{LOVV}$, and $\ml{Airport} \preceq_{\ml{Topic}} \ml{Infrastructure}$.
The cell $c_2$ therefore covers $c_1$---and, by the same argument, all four sector-level cells of the $\ml{25\text{-}Aug}$ slice in Figure~\ref{fig:cube-example}; every cell of the cube is covered by $c_\top$.
In contrast, the vector $\mathbf{d}_3 = \{\ml{Time} := \ml{24\text{-}Aug},\, \ml{Location} := \ml{Wien\text{-}N},\, \ml{Topic} := \ml{Airport}\}$ is not covered by $\mathbf{d}_2$, because $\ml{24\text{-}Aug} \not\preceq_{\ml{Time}} \ml{25\text{-}Aug}$: knowledge about the previous day lies outside the scope of $c_2$.\EndEx
\end{example}

The knowledge represented in each cell is then expressed in
a language $\Lcal_\Sigma$, called the \emph{object language},
which in turn is based on an \emph{object vocabulary}
$\Sigma = (\mathrm{NR}_\Sigma, \mathrm{NC}_\Sigma, \mathrm{NI}_\Sigma)$.

We provide a simplified definition of KG-OLAP cube from \cite{schuetz2021}, showing a clearer association
of cells to their object knowledge contents:

\begin{definition}[KG-OLAP cube]
A \emph{KG-OLAP cube} $\CKB$ over a cube vocabulary $\Omega$ and an object
vocabulary $\Sigma$ is a structure
\[
  \CKB = \stru{\Gl, K_F}
\]
such that:
\begin{itemize}
\item
  the upper layer
  $\Gl = \stru{\preceq, \Ccal, \Tcal_\Sigma}$ consists of
  the coverage relation $\preceq$, induced by the member orderings $\preceq_A$ for all $A \in D$
  and the cell association mapping $\cn$,
  a mapping
$\mathcal{C} : F \rightarrow K_F$
that assigns to each cell name $c \in F$ its corresponding knowledge module $K_c$ and 
a global terminological (TBox) component $\Tcal_\Sigma \subseteq \Lcal_\Sigma$
expressing cell-independent axioms;
\item 
  the lower layer is defined by
  $K_F = \{K_c\}_{c \in F}$, where every $K_c \subseteq \Lcal_\Sigma$ is a local assertional (ABox) knowledge module
  for a cell name $c \in F$.
\end{itemize}
\end{definition}

\noindent
Note that, with respect to the original definition of KG-OLAP cube,
the current definition highlights the contextual structure in $\Gl$
(now defined in terms of coverage relations and module mapping) and
considers the more specific case in which cells are associated
with a single module containing ABox axioms.
Figure~\ref{fig:cube-example} illustrates these notions with a cube from the ATM use case of Section~\ref{sec:case}: three dimensions with three-level hierarchies, four sector-level cells, and two covering contexts.
Examples \ref{ex:cells}--\ref{ex:propagation} explain the cube.

\begin{example}[KG-OLAP cube]
\label{ex:cube}
The cube of Figure~\ref{fig:cube-example} assigns to the cell $c_1$ of Example~\ref{ex:cells} the local knowledge module $K_{c_1} = \{\role{avail}(\indiv{LNZ}, \indiv{CLOSED})\}$, recording the closure of Linz airport ($\indiv{LNZ}$) on 25~August in sector Wien-N.
The sibling cells of the same slice record that Vienna airport ($\indiv{VIE}$) remains open, $\role{avail}(\indiv{VIE}, \indiv{OPEN})$, that the navigation aid $\indiv{FRE}$ transmits on $\role{freq}(\indiv{FRE}, \reified{\text{113.5~MHz}})$, and that the navigation aid $\indiv{FMD}$ is out of service, $\role{status}(\indiv{FMD}, \indiv{INTERRUPT})$.
The covering cell $c_2$ holds knowledge that pertains to the infrastructure of the entire flight information region on that day rather than to a single sector. Flights bound for the closed airport are rerouted, and $\indiv{VIE}$ is temporarily served by $\indiv{FRE}$ while $\indiv{FMD}$ is interrupted:
\begin{align*}
K_{c_2} = \{&\role{flightsReroutedTo}(\indiv{LNZ}, \indiv{VIE}),\\
            &\role{servedBy}(\indiv{VIE}, \indiv{FRE})\}.
\end{align*}
The most general cell holds reference knowledge:
\begin{align*}
K_{c_\top} = \{&\concept{Airport}(\indiv{LNZ}),\ \concept{Airport}(\indiv{VIE}),\\
               &\concept{VOR\text{-}DME}(\indiv{FRE}),\ \concept{VOR\text{-}DME}(\indiv{FMD}),\\
               &\role{servedBy}(\indiv{LNZ}, \indiv{FRE}),\ \role{servedBy}(\indiv{VIE}, \indiv{FMD})\}.
\end{align*}
The global TBox $\Tcal_\Sigma$, shared by all cells, may state, for example, that flights are rerouted only between airports:
\begin{align*}
\exists\role{flightsReroutedTo}.\top &\isa \concept{Airport},\\
\top &\isa \forall\role{flightsReroutedTo}.\concept{Airport}.
\end{align*}\EndEx
\end{example}

We can formalize the DL semantics of this structure
(and thus, its intended reading)
by adapting the definitions of interpretation
and model from the original KG-OLAP formalization~\cite{schuetz2021}.

\begin{definition}[Interpretation]
 An \emph{interpretation} for $\stru{\Omega, \Sigma}$
 is a structure $\IC = \stru{\Mcal, \I}$ such that:
  \begin{enumerate}[label=(\roman*).]
	\item
	  $\Mcal$ is a DL interpretation of $\Omega \cup \Sigma$;
	\item
		for every $c\in F$, $\I(c)$ is a DL interpretation over $\Sigma$ such that
		$\Delta^{\I(c)} = \Delta^{\Mcal}$ and,
        for $a \in \mathrm{NI}_\Sigma$, $a^{\I(c)} = a^{\Mcal}$.
  \end{enumerate}
\end{definition}

\begin{definition}[KG-OLAP cube model]
  An interpretation $\IC = \stru{\Mcal, \I}$ is a 
  \emph{KG-OLAP cube model} of a cube $\CKB$
  iff the following conditions hold:
\begin{enumerate}[label=(\roman*).]
\item
  for $\alpha \in \Tcal_\Sigma$, $\Mcal \models \alpha$
  (satisfaction of global TBox);
\item
  for every $c \in F$, $\I(c) \models K_c$
  (satisfaction of local ABoxes);
\item 
  for $\alpha \in \Tcal_\Sigma$ and $c \in F$, $\I(c) \models \alpha$
  (local satisfaction of TBox axioms);
\item
  if $c_1, c_2 \in F$ with $c_1 \prec c_2$ in $\Gl$
  and $\Ccal(c_2) = K_{c_2}$, then
  $\I(c_1) \models K_{c_2}$
  (propagation of knowledge modules).
\end{enumerate}
\end{definition}

We note that, considering the restricted
definition of cube provided above, this definition
of the semantics is also simplified with respect
to the more general one provided in \cite{schuetz2021} 
(e.g., we associate only a module for each cell and 
we do not consider context classes).

The KG-OLAP cube separates context-specific assertional knowledge
(local modules $K_c$) from global terminological knowledge
$\mathcal{T}_\Sigma$. 
This separation enables reasoning across contexts while preserving the multidimensional structure of the data.
The local knowledge modules $K_c$ are expressed in an object language $\mathcal{L}_\Sigma$:
in the original formulation of the KG-OLAP framework~\cite{schuetz2021}, the object language is the $\mathcal{SROIQ}$-RL description logic~\cite{BozzatoES:18} over the vocabulary $\Sigma$.

By the definition of the semantics, note that 
the KG-OLAP model also supports \textit{knowledge propagation}: if $c_1 \preceq c_2$, the assertions in the more general context $c_2$ are inherited by the more specific context $c_1$.
Knowledge propagation enables reasoning across different levels of granularity of the knowledge.

\begin{example}[Knowledge propagation]
\label{ex:propagation}
In the cube of Figure~\ref{fig:cube-example}, $c_1 \preceq c_2 \preceq c_\top$, and thus every cube model satisfies $K_{c_2}$ and $K_{c_\top}$ also at $c_1$.
Within $c_1$, the local closure assertion therefore combines with the inherited knowledge: $\indiv{LNZ}$ is a closed airport whose flights are rerouted to the airport $\indiv{VIE}$---although no single module asserts all of these facts together.
Propagation follows the coverage relation only: for instance, the frequency of $\indiv{FRE}$, asserted in the sibling cell for topic $\ml{Navaid}$, is not visible in $c_1$.\EndEx
\end{example}

Section~\ref{sec:architecture} builds on this formal definition and describes the KG Lakehouse
architecture tier by tier.
\section{Architecture}
\label{sec:architecture}

The KG Lakehouse manages contextualized knowledge graphs under the \emph{Three Vs} of big data~\cite{Laney2001}: volume, velocity, and variety.
Data arrive continuously, in the form of documents of heterogeneous formats, accumulating to a volume that exceeds what any individual analytical task typically draws on: a task often needs only the knowledge relevant for the context, not the entire enterprise KG.
From this observation, two requirements follow. 
Documents must be ingested as they arrive, before the tasks that will use the documents are known, meaning that the ingestion of a document cannot wait for the task using the document. 
Ultimately, a query must be answered for the knowledge relevant for the contexts of interest for a particular task, the knowledge extracted from the documents available at query time, and recently arriving documents are not withheld from the answer because the system must wait for a monolithic KG to be rebuilt.

The KG Lakehouse architecture meets adapts and extends the data lakehouse paradigm~\cite{Armbrust2021} for the management of CKGs.
A data lakehouse retains raw data while imposing the management structure of a data warehouse: metadata, indexes, and caching.
Similarly, the KG Lakehouse retains the arriving documents unchanged and imposes a contextual index: each document is recorded under the contexts which the document's contents are valid for.
When a query names the required contexts, the KG comprising the knowledge for those contexts is constructed on demand from the indexed documents rather than extracted from a materialized monolithic KG.
Where a data lake defers interpretation of the documents to the moment of reading (\emph{schema on read}), the KG Lakehouse determines the context on arrival while deferring interpretation to the query.
Constructed KGs may be cached for reuse, but what the system holds durably is the documents and the index.

The multidimensional context model of Section~\ref{sec:kgolap} supplies the structure for the constructed KGs.
Dimensions with hierarchically organized members span the space of contexts, each context is a cell of the cube and carries its own local knowledge module, and the coverage relation orders contexts from the specific to the general.
The index associates documents with cells, construction realizes the mapping from cells to modules, and the contextual operations are defined over the same notions.

Figure~\ref{fig:conceptual_architecture} shows the resulting organization: five tiers, ordered along the path from an arriving document to a delivered CKG.
The first two tiers handle the incoming documents, the remaining three tiers handle the work related to processing a query.
\begin{itemize}
    \item The \emph{Ingestion} tier receives documents of heterogeneous formats and determines the contexts which the documents' contents are valid for.

    \item The \emph{Data Lake} tier retains the documents and maintains the contextual index that records association between documents and contexts.

    \item The \emph{Knowledge Graph} tier constructs, for the contexts a query addresses, the local knowledge modules those contexts hold.

    \item The \emph{OLAP} tier selects and combines contexts, and renders the resulting cube into the representation that the query requested.

    \item The \emph{Frontend} tier receives queries and delivers the constructed CKG to the applications that consume the CKG, among them applications for reporting, data mining, and rule-based reasoning.
\end{itemize}

The decomposition into tiers allows for the modularity, elasticity, and scalability required of a cloud-native architecture able to manage CKGs at scale.
Each tier is delimited by what it commits to, that is, by what it undertakes to provide the other tiers, and not by the data it holds.
Of the five, only the Data Lake Tier holds data durably, and it too is delimited by its commitment rather than by that holding: the tiers divide the work; they do not layer the storage.
The other tiers depend on that commitment alone, and not on how a tier meets it, so a tier can be extended, or realized by other means, without disturbing them.
Because the durable state is the documents and the index, constructing a graph carries no state of its own, and the capacity devoted to construction can therefore grow and shrink with demand.
The index decouples the two sides of the path, so the capacity for ingesting documents and the capacity for answering queries are adjusted independently.
Contexts partition the knowledge, and the work for distinct contexts is independent, which allows parallel processing of the contexts a query addresses.

The proposed KG Lakehouse architecture is independent of the application domain.
Domain-specific differences in the data are configurable parameters, including the cube schema, the dimensions, levels, and member hierarchies a client registers (Section~\ref{sec:frontend}), a content analyzer per source format on the arrival side to determine relevant contexts from the documents (Section~\ref{sec:backend_tier}), and a semantic-lifting function per source format on the query side (Section~\ref{sec:kg_tier}) to convert the raw documents into knowledge statements.
The tiers themselves are defined over the multidimensional context model.
Another source format is accommodated by an additional analyzer and lifter, and another domain by an additional cube schema.
Air traffic management is the setting of the running example and of the evaluation, but the architecture is not limited to that domain.

Each of the following subsections presents and formalizes a particular tier of the architecture.
How a running system can be built based on this architecture is the subject of Section~\ref{sec:poc}.

\subsection{Ingestion Tier}
\label{sec:backend_tier}
The \emph{Ingestion Tier} receives arriving documents and extracts the contextual metadata that assigns their content to the contexts in which it is valid.
The ingestion tier derives no knowledge and builds no graph. 
Ingestion records the assignment of document to contexts, and defers the extraction of knowledge, the question \emph{what} the content means; the interpretation happens at query time (Section~\ref{sec:kg_tier}). 
The work done when a document arrives is therefore independent of how the data will later be queried.

A document is stored unchanged and remains the authoritative source of its own content.
The analysis is separated from storage, so the rate at which documents arrive does not constrain the rate at which the data can be analyzed, and analysis capacity can be raised independently of the ingestion (Section~\ref{sec:evaluation}).

Analysis is format-specific. 
A content analyzer reads documents of one source format, so a further format is accommodated by a further analyzer rather than by a change to the tier.
We illustrate the approach with an AIXM content analyzer, which processes documents describing events in the air traffic management (ATM) network.

The extracted metadata describes the context in which the data is valid and is multidimensional, typically comprising temporal, spatial, and thematic dimensions, the thematic dimension covering topics such as airspace closures or navigation aid restrictions.
Each dimension is organized hierarchically (e.g., day, month, and year).
A document may hold several raw data items, and the analyzer treats each of them separately.

\begin{definition}[Raw data item]
Let $\mathcal{R}_{\mathcal{S}}$ denote the set of raw data items encoded in various formats (e.g., XML, JSON, CSV) from a set $\mathcal{S}$ of source-format schemas, such as the XML schema of AIXM.
\end{definition}

\begin{example}[Raw data item]\label{ex:raw_data_item}
Consider the following simplified AIXM XML fragment describing the closure of Linz airport in the scenario of Figure~\ref{fig:cube-example}:

\small
\begin{verbatim}
<AirportHeliport>
  <timeSlice>
    <validTime>25-Aug</validTime>
    <designator>LNZ</designator>
    <availability>CLOSED</availability>
  </timeSlice>
</AirportHeliport>
\end{verbatim}
\normalsize
This XML fragment corresponds to a raw data item $x_1 \in \mathcal{R}_{\mathcal{S}}$ extracted from an ingested AIXM document $doc_1$.\EndEx
\end{example}

\begin{definition}[Extraction of raw data items from documents]
Let $\Doc$ denote the set of ingested documents.
An extraction function
\[
\phi : \Doc \rightarrow 2^{\mathcal{R}_{\mathcal{S}}}
\]
maps each document to a finite set of raw data items.
Each raw data item $x \in \mathcal{R}_{\mathcal{S}}$ represents a logical unit of information extracted from a document.
\end{definition}

Each raw data item $x \in \mathcal{R}_{\mathcal{S}}$ is processed by a content analyzer, which extracts the dimensional vectors that position the item within the multidimensional space.

\begin{definition}[Context assignment]
The mapping of raw data items to sets of points in the multidimensional space is defined by a function
\[
\kappa : \mathcal{R}_{\mathcal{S}} \rightarrow 2^{\Dimspace},
\]
which extracts a set of dimensional vectors $\kappa(x)$ from a raw data item $x$.
\end{definition}
\noindent
Each vector $\mathbf{d} \in \kappa(x)$ defines a cell $\cn(\mathbf{d})$ to which the item is assigned; an item whose validity spans several members of a dimension yields one vector per covered member---one item, many cells.

\begin{example}[Context assignment]
\label{ex:context_assignment}
For the raw data item $x_1 \in \phi(doc_1)$ of Example~\ref{ex:raw_data_item}, the content analyzer reads three values from the fragment. The valid time $\ml{25\text{-}Aug}$ gives the day, the aerodrome designator $\ml{LNZ}$ gives the sector $\ml{Wien\text{-}N}$, and the element type \texttt{AirportHeliport} gives the topic $\ml{Airport}$. The closure is valid on a single day, so $\kappa(x_1)$ contains exactly one dimensional vector, the $\mathbf{d}_1$ of Example~\ref{ex:cells}.
The item is thus assigned to the cell $c_1 = \cn(\mathbf{d}_1)$ of Figure~\ref{fig:cube-example}.\EndEx
\end{example}

The extracted contexts populate the contextual index of Section~\ref{sec:datalake_tier}, which associates each document with its cells, so that construction reads only the documents of the queried cells rather than the whole corpus.

\subsection{Data Lake Tier}
\label{sec:datalake_tier}

The \emph{Data Lake Tier} manages the storage and contextual indexing of the ingested documents. 
This tier distinguishes between the physical storage of raw documents and the logical representation of raw data items extracted from these documents. This separation lets document storage and the index scale
independently, and keeps the index small relative to the corpus
(Section~\ref{sec:evaluation}), so that a context lookup never scans the documents.

The contextual index captures the association between cells and documents based on the contexts of the raw data items contained in the documents.

\begin{definition}[Contextual index]
Let $\phi$ and $\kappa$ be the extraction and context assignment functions of Section~\ref{sec:backend_tier}, and $\cn$ the mapping from coordinates to cells of Section~\ref{sec:kgolap}.
The \emph{contextual index} is then defined as a relation
\[
\Idx \subseteq F \times \Doc
\]
such that:
\[
(c, doc) \in \Idx
\;\;\Longleftrightarrow\;\;
\exists x \in \phi(doc),\ \exists \mathbf{d} \in \kappa(x): \cn(\mathbf{d}) = c.
\]
\end{definition}

Context assignment is performed at the level of raw data items, while the index is maintained at the level of documents.
Since a document may contain multiple data items associated with different contexts, and each item may be assigned to multiple cells, a document may be linked to multiple cells.
The index therefore captures a coarse-grained association between documents and contexts, while fine-grained associations are recovered at query time.

\begin{example}[Contextual index]
\label{ex:contextual_index}
Consider the document $doc_1$ of Examples \ref{ex:raw_data_item} and \ref{ex:context_assignment}, with $\phi(doc_1) = \{x_1\}$, and a second AIXM document $doc_2$ with two raw data items, $\phi(doc_2) = \{x_2, x_3\}$: $x_2$ reports the interruption of the navigation aid $\ml{FMD}$ and $x_3$ reports that Vienna airport remains open, both on $\ml{25\text{-}Aug}$ (Figure~\ref{fig:cube-example}).
Context assignment places $x_2$ in the cell $c_4$, whose coordinates differ from $\mathbf{d}_1$ in $\ml{Location} := \ml{Wien\text{-}E}$ and $\ml{Topic} := \ml{Navaid}$, and $x_3$ in the cell $c_3$, differing from $\mathbf{d}_1$ only in $\ml{Location} := \ml{Wien\text{-}E}$.
Restricted to these two documents, the contextual index is
\[
\Idx =
\{
(c_1, doc_1),\
(c_4, doc_2),\
(c_3, doc_2)
\}:
\]
the document $doc_2$ is linked to two cells because its items belong to different contexts.\EndEx
\end{example}

Persisting the relation $\Idx$ materializes the association between cells and documents, and thereby separates a document's physical location in storage from its logical positions in the KG-OLAP cube.

At query time, the set $RD_c$ of relevant raw data items for a context $c$ is derived by retrieving all documents associated with $c$ and extracting their raw data items again with $\phi$:
\[
RD_c =
\bigcup_{\{doc \in \Doc \,\mid\, (c, doc) \in \Idx\}} \phi(doc), 
\]
where $\phi(doc)$ denotes the set of raw data items contained in document $doc$.

Retrieval therefore reads only the documents of the queried cells, and recovers their items by applying $\phi$ again rather than by storing the items themselves.

\subsection{Knowledge Graph Tier}
\label{sec:kg_tier}

The \emph{Knowledge Graph Tier} is responsible for the transition from indexed raw data to a structured, contextualized knowledge graph.
Rather than building graphs when documents arrive, it constructs local knowledge modules on demand, when a query names the contexts it needs.
A constructed graph is not part of the durable state of the system. It is delivered to the client that requested it, while the documents and the index remain what the system maintains. Constructed modules may be retained for reuse (Section~\ref{sec:caching}), without becoming authoritative.
The Knowledge Graph Tier performs two functions. \emph{Resolution} extracts the scope and the level assignment from a KG specification, and retrieves the documents that the index associates with the selected cells (Section~\ref{sec:olap}), yielding the raw data items $RD_c$.
\emph{Construction} lifts those items into logical assertions and aggregates them into the local knowledge modules of the cube.

\subsubsection{Cube Construction}

The construction process retrieves relevant raw data items for specific contexts on demand, 
based on a user-defined KG specification, and constructs a contextualized knowledge graph (CKG) 
in the form of a KG-OLAP cube.

Construction begins with a KG specification. 
This specification defines the applicability of the desired CKG along the dimensions of the cube schema, in the ATM use case, its temporal, spatial, and thematic scope.

\begin{definition}[KG specification]
A \emph{KG specification} names a cube schema and assigns to each dimension $A \in D$ at most one member $e_A \in D_A$, at any level of the dimension's hierarchy, and optionally a target level $l_A \in L_A$.
From the specification we extract the \emph{scope vector} $\mathbf{e} \in \Dimspace$, obtained by completing the named members with the root $all_A$ of every unconstrained dimension, and the \emph{level assignment} $\mathbf{l}$.
\end{definition}

The scope $\mathbf{e}$ and the level assignment $\mathbf{l}$ are the parameters of the contextual operations, which define from them the cells of the result and the granularity at which their knowledge is delivered (Section~\ref{sec:olap}). Whether reasoning is applied and in which representation the result is rendered accompany the specification but do not affect context selection.
For example, a request for $\ml{AirportHeliport}$ data for Austrian airspace during January 2018 constrains the time dimension to the member $\ml{January~2018}$ at the month level (its ancestor, the year 2018, follows from the hierarchy), the location to the $\ml{Austria}$ member at the territory level, and the topic to the $\ml{AirportHeliport}$ member at the category level.

Resolution parses the specification into the scope $\mathbf{e}$ and the level assignment $\mathbf{l}$.
For each selected cell it then retrieves the identifiers of the documents that $\Idx$ associates with that cell.

The raw data items associated with a context $c \in F$ form the set $RD_c$ introduced in Section~\ref{sec:datalake_tier}, i.e., the items extracted by $\phi$ from the documents that the index $\Idx$ associates with $c$.
The selected cells are not only those the scope covers. They include the covering contexts (e.g., national regulations) whose knowledge the covered cells inherit through the \textit{context coverage} relation ($\preceq$).
Each selected cell, covered or covering, enters the construction with its own set $RD_c$. The covering modules reach the cells they cover only at delivery, through knowledge propagation (Section~\ref{sec:olap}).
Propagation never adds items to $RD_c$.

The corresponding documents are retrieved from the lake and their content is extracted by $\phi$, obtaining the raw data items. Semantic lifting $\lambda$ (Definition~\ref{def:lifting}) turns each item into logical assertions, which are aggregated into the local knowledge modules $K_c$ (Definition~\ref{def:construction}).

This design instantiates the formal KG-OLAP model \cite{schuetz2021} by materializing local knowledge modules from indexed data, where hierarchical dimensions ensure the correctness of OLAP operations.
This process avoids full knowledge graph materialization by constructing
only the context-relevant fragments required for a given query.

\subsubsection{Extraction of Assertional Knowledge}

The transition from indexed raw data to formal graph assertions is
performed by resolving the index $\Idx$ and applying semantic
lifting to the raw data items derived from the indexed documents.
Semantic lifting refers to the transformation of raw data items into a formal, machine-interpretable representation by mapping them to logical assertions defined over a given vocabulary.

Let $\Idx \subseteq F \times \Doc$ denote the
contextual index produced during ingestion.
For a specific cell $c \in F$, the associated set of raw data items $RD_c$ is used as input for semantic lifting.

\begin{definition}[Semantic lifting]
\label{def:lifting}
A \emph{semantic lifting function} $\lambda_{\mathcal{S}, \Sigma}$ transforms each raw data item 
of format $\Scal$
into a set of logical assertions of the object language $\Lcal_\Sigma$
\[
\lambda_{\mathcal{S}, \Sigma} : \mathcal{R}_{\mathcal{S}} \rightarrow 2^{\mathcal{L}_\Sigma}
\]
%
%
\end{definition}%
\noindent
In the following, we write $\lambda$ for $\lambda_{\mathcal{S}, \Sigma}$ whenever the parameters are understood.

\begin{example}[Semantic lifting]
\label{ex:lifting}
Consider the cube of Figure~\ref{fig:cube-example} and the two ingested AIXM documents of Example~\ref{ex:contextual_index}: $doc_1$, whose raw data item $x_1$ states that Linz airport is closed on 25~August, and $doc_2$, whose item $x_2$ states that the navigation aid $\indiv{FMD}$ is out of service on the same day.
Semantic lifting translates each item's content to ABox assertions in the object language of the cube:
\[
\lambda(x_1) = \{ \role{avail}(\indiv{LNZ}, \indiv{CLOSED}) \},
\]
\[
\lambda(x_2) = \{ \role{status}(\indiv{FMD}, \indiv{INTERRUPT}) \}.
\]
The lifting $\lambda$ operates on the content of an item alone. The cube plays no role in the transformation, and each item lifts to the same assertions under whichever cell it is indexed.
The association with contexts is established separately, by the contextual index $\Idx$: during ingestion, the time, location, and topic extracted from the same content yielded $(c_1, doc_1) \in \Idx$ for the cell $c_1$ of Example~\ref{ex:cells} and $(c_4, doc_2) \in \Idx$ for the sibling cell $c_4$ with $\ml{Topic} := \ml{Navaid}$.
Resolving the index at query time gives $RD_{c_1} = \phi(doc_1) = \{x_1\}$, and aggregation yields the local module $K_{c_1} = \lambda(x_1) = \{\role{avail}(\indiv{LNZ}, \indiv{CLOSED})\}$ of Example~\ref{ex:cube}.\EndEx
\end{example}

\begin{definition}[Knowledge Module Construction]
\label{def:construction}
The \textit{local knowledge module} $K_c$ for a cell $c \in F$ is defined as:
\[
K_c = \bigcup_{x \in RD_c} \lambda(x)
\]
that is, the union of the assertion sets lifted from the raw data items associated with
the cell $c$. 
\end{definition}
The resulting KG-OLAP cube is the collection of the pairs
$\langle c, K_c \rangle$ for the selected cells, representing the assertional
base (ABox) of the knowledge graph.

\begin{example}[Knowledge module construction]
\label{ex:construction}
Continuing Example~\ref{ex:lifting}, suppose a third AIXM document $doc_3 \in \Doc$ announces the operational consequences of the closure for the entire flight information region and contains two raw data items, $\phi(doc_3) = \{x_4, x_5\}$: $x_4$ states that flights bound for Linz are rerouted to Vienna, and $x_5$ states that $\indiv{VIE}$ is temporarily served by the navigation aid $\indiv{FRE}$.
Ingestion indexed the document under the cell $c_2$ of Example~\ref{ex:cells}, that is, $(c_2, doc_3) \in \Idx$, and thus $RD_{c_2} = \phi(doc_3) = \{x_4, x_5\}$.
The local knowledge module aggregates the lifted fragments of both items,
\[
K_{c_2} = \lambda(x_4) \cup \lambda(x_5),
\]
yielding exactly the module $K_{c_2}$ of Example~\ref{ex:cube}.\EndEx
\end{example}

\subsubsection{Integration of Terminological Knowledge}

While $K_c$ captures the assertional knowledge (ABox) derived from raw data for a specific context, the global TBox $\Tcal_\Sigma$ governs the interpretation of the knowledge, the terminological component of the upper layer $\Gl$ (Section~\ref{sec:kgolap}). 
Shared across all cells, the TBox comprises, for example, axioms of domain-specific ontologies.
By the cube-model semantics, the axioms of $\Tcal_\Sigma$ hold in every cell.
Reasoning within a cell therefore evaluates the extended module
\[
M_c = K_c \cup \Tcal_\Sigma.
\]
When a cell is delivered with its inherited knowledge, the propagated module $K^{\preceq}_c$ (Section~\ref{sec:olap}) takes the place of $K_c$.
Construction itself never materializes $\Tcal_\Sigma$ into the modules. 
$K_c$ remains purely assertional, and the TBox enters only when reasoning is requested.
The extended module provides the logical basis for inferring implicit knowledge within each context.

\subsubsection{Caching}
\label{sec:caching}

On-demand construction concentrates its cost in semantic lifting, and successive queries overlap. Two analyses of the same region or period address many of the same cells.
The system therefore caches constructed knowledge modules.
The unit of caching is the formal unit of the model. An entry is a pair $\langle c, K_c \rangle$, the constructed module of a single cell, serialized in one of the output representations and keyed by the cell identifier $\cn(\mathbf{d})$ that the contextual index also uses.
The cache is thus a partial materialization of the cube mapping $\Ccal: F \rightarrow K_F$ (see Section~\ref{sec:kgolap}), covering the recently queried cells.

Correctness follows from the shape of the construction function.
By definition, $K_c = \bigcup_{x \in RD_c} \lambda(x)$, hence the module is determined solely by the raw data items $RD_c$, which the index determines, and by the lifting $\lambda$, which depends only on the content of each item, never on the query.
Construction is therefore deterministic, and a cache hit is indistinguishable from a fresh construction. 
The cache can save time but cannot change a result.
The same determinism yields a precise validity condition: an entry $\langle c, K_c \rangle$ remains correct exactly as long as the index restricted to $c$ is unchanged.
The condition is stable because stored documents are immutable. 
The store keeps each document exactly as it arrived, so the index is the only input to construction that changes, and the index changes only through the ingestion pipeline.

An \emph{invalidation} mechanism enforces the validity condition with information the ingestion pipeline already possesses.
Indexing a document computes the set of cells the $RD_c$ of which the document extends, by the same many-to-many document--context assignment that fills the index. 
The pipeline announces precisely these cell identifiers, and the affected modules in the cache are discarded.
Invalidation is exact: an arriving document removes the modules from the cache that document invalidates, and no others.
Since a module can always be reconstructed from the indexed documents, a failing cache costs time, but never correctness.

Caching at cell granularity, rather than at the granularity of query results, allows cached modules to feed the contextual operations of the OLAP Tier (Section~\ref{sec:olap}).
The delivered modules are unions of local knowledge modules. 
Coverage extends a cell by its covering cells, and roll-up merges the cells that coarsen to the same coordinates. 
A union assembled from cached modules is identical to one assembled from freshly constructed modules.
Consequently, two queries need not coincide to share work. 
Any overlap in cells is reused, whatever their scopes or target granularities.
\subsection{OLAP Tier}\label{sec:olap}

The \emph{OLAP Tier} operates on the contextualized knowledge graph as a multidimensional cube.
Whereas the Knowledge Graph Tier constructs the local knowledge modules $K_c$, the OLAP Tier determines \emph{which} cells a query addresses and \emph{at which granularity} their knowledge is delivered, and renders the result in a representation suited to the consuming application.
It comprises two functions. The \emph{contextual operations} select and combine cells. Slice and dice selects the cells the requested scope addresses, and roll-up merges the modules of covered cells into coarser contexts. The \emph{knowledge graph transformation} then serializes the resulting cube as RDF quads, as a labeled property graph, or as a GraphFrame for distributed processing.
The following subsections formalize both.

\subsubsection{Contextual Operations}

Contextual operations manipulate the cube at the level of its cells, leaving the contents of the local knowledge modules unchanged.
Slice and dice restricts the multidimensional space to the cells whose coordinates are comparable with the requested coordinates in every dimension. These are the cells the scope covers, together with the cells coarser in at least one dimension whose knowledge they inherit. Roll-up coarsens the granularity by mapping each covered cell to its ancestor at the target level and merging the modules of cells that collapse to the same coordinates, the union variant of the KG-OLAP merge operation~\cite{schuetz2021}.
In the KG Lakehouse, these operations are not applied to a standing cube. The system plans them over the contextual index and constructs the merged modules directly.
The KG specification introduced in Section~\ref{sec:kg_tier} supplies the two parameters of these operations: the requested coordinates (the scope) and, optionally, a target level per dimension.
Throughout, we identify a cube with its cell-to-module mapping.

\begin{definition}[Slice and dice]
Let $\mathbf{e} \in \Dimspace$ be the requested scope. Slice and dice restricts the cube $\mathcal{C}$ to the cells whose coordinates are comparable with the scope in every dimension,
\[
F_{\mathbf{e}} = \{ \cn(\mathbf{d}) \mid \mathbf{d} \in \Dimspace,\ \forall A \in D: d_A \preceq_A e_A \;\text{or}\;\ e_A \preceq_A d_A \}
\]
yielding a new cube, the subcube $\mathcal{C}|_{F_{\mathbf{e}}} : F_{\mathbf{e}} \rightarrow 2^{\mathcal{L}_\Sigma}$, which assigns to each cell $c \in F_{\mathbf{e}}$ its local knowledge module $K_c$.
\end{definition}

Comparability is checked per dimension, not on the dimensional vectors as wholes.
$F_{\mathbf{e}}$ therefore contains, besides the cells that the scope covers ($c \preceq \cn(\mathbf{e})$), every cell that is coarser than the scope in at least one dimension, including a cell coarser in one dimension and finer in another, which covers part of the scope without being comparable with the scope vector itself. Dropping such a cell would cut the covered cells off from knowledge they inherit. By knowledge propagation (Section~\ref{sec:kgolap}), a cell inherits the assertions of every cell covering it.
Because the dimension hierarchies are trees, every cell covering a cell of $F_{\mathbf{e}}$ lies itself in $F_{\mathbf{e}}$. The subcube is closed under coverage, so the restriction loses no inherited knowledge.
The module delivered for a covered cell extends its local module with the inherited assertions.

\begin{definition}[Propagated knowledge module]
For a cell $c \in F$, the propagated knowledge module is
\[
K^{\preceq}_c = \bigcup_{\{c' \in F \,\mid\, c \preceq c'\}} K_{c'},
\]
the union of the local module of $c$ with the local modules of all cells covering $c$.
\end{definition}

Roll-up is guided by the level assignment $\mathbf{l}$ of the KG specification (Section~\ref{sec:kg_tier}). Its definition rests on two auxiliary notions.

\begin{definition}[Ancestor at a level]
For a member $d_A \in D_A$ and a level $l_A \in L_A$, the \emph{ancestor} of $d_A$ at $l_A$ is
\[
anc_{l_A}(d_A) =
\begin{cases}
e_A & \text{if } d_A \preceq_A e_A \text{ and } lev(e_A) = l_A,\\
d_A & \text{if no such } e_A \text{ exists}.
\end{cases}
\]
\end{definition}

The ancestor is unique because $\prec_A$ orders $D_A$ in a tree.

\begin{definition}[Coordinate coarsening]
Given a level assignment $\mathbf{l}$, the \emph{coordinate coarsening} $\rho_{\mathbf{l}}: \Dimspace \to \Dimspace$ applies the ancestor per dimension: $\rho_{\mathbf{l}}(\mathbf{d}) = \{A := anc_{l_A}(d_A) \mid A \in D\}$.
\end{definition}

\begin{definition}[Roll-up]
Given the subcube $\mathcal{C}|_{F_{\mathbf{e}}}$ and a level assignment $\mathbf{l}$, the \emph{roll-up} is the cube $\mathcal{C}' : F' \rightarrow 2^{\mathcal{L}_\Sigma}$ such that:
\begin{itemize}
\item $F' = \{\, \cn(\rho_{\mathbf{l}}(\cn^-(c))) \mid c \in F_{\mathbf{e}},\ c \preceq \cn(\mathbf{e}) \,\}$ (the coarsened cells);
\item for each $c' \in F'$, the merged module is
\[
K'_{c'} = \bigcup_{c \,\in\, S_{c'}} K^{\preceq}_c,
\]
where $S_{c'} = \{ c \in F_{\mathbf{e}} \mid c \preceq \cn(\mathbf{e}) \wedge \rho_{\mathbf{l}}(\cn^-(c)) = \cn^-(c') \}$ is the set of covered cells that coarsen to $c'$.
\end{itemize}
\end{definition}

The covering cells of $F_{\mathbf{e}}$ contribute to the merged modules through the propagated modules $K^{\preceq}_c$ rather than as targets of their own.
The contextual operations leave the terminological component unchanged, and the coverage relation restricts to the retained cells (for roll-up, the relation induced on the coarsened cells), so each result is again a cube $\stru{\Gl', K_{F'}}$ in the sense of Section~\ref{sec:kgolap}. Both operations therefore compose, and the knowledge graph transformation (Section~\ref{sec:transformation}) consumes their result uniformly.

\begin{example}[Roll-up]
\label{ex:rollup}
In the cube of Figure~\ref{fig:cube-example}, consider the two airport cells of the $\ml{25\text{-}Aug}$ slice: the cell $c_1$ of Example~\ref{ex:cells}, with $K_{c_1} = \{\role{avail}(\indiv{LNZ}, \indiv{CLOSED})\}$, and its sibling $c_3$, whose coordinates differ from $\mathbf{d}_1$ only in $\ml{Location} := \ml{Wien\text{-}E}$, with $K_{c_3} = \{\role{avail}(\indiv{VIE}, \indiv{OPEN})\}$ (Example~\ref{ex:cube}).
Rolling up the $\ml{Location}$ dimension to level $\ml{FIR}$ replaces both location coordinates by their ancestor $\ml{LOVV}$: the two cells collapse into a single target cell $c'$, whose merged module unites their propagated modules,
\[
K'_{c'} = K^{\preceq}_{c_1} \cup K^{\preceq}_{c_3}.
\]
The local contributions deliver the availability of both airports at FIR granularity; through the propagated modules, $c'$ also carries the knowledge inherited from the covering cells $c_2$ and $c_\top$ of Example~\ref{ex:cube}, such as the rerouting of flights from the closed $\indiv{LNZ}$ to $\indiv{VIE}$.\EndEx
\end{example}

The KG Lakehouse does not first build the cube at source granularity and then merge the cells.
Rather, the KG Lakehouse computes each source cell's final coordinates up front and constructs the merged modules directly from the raw data items (Section~\ref{sec:kg_tier}).
At the assertional level, the two orders coincide: merging the constructed modules equals constructing one module from the merged raw data items.

\begin{proposition}[Construction-time merge]\label{prop:merge}
For every target cell $c'$,
\[
\bigcup_{c \,\in\, S_{c'}} K_c
\;=\;
\bigcup_{x \,\in\, RD_{S_{c'}}} \lambda(x),
\qquad
RD_{S_{c'}} = \bigcup_{c \,\in\, S_{c'}} RD_c,
\]
that is, merging the constructed modules of the collapsing cells equals constructing one module from their combined raw data items.
\end{proposition}

\begin{proof}
Immediate from $K_c = \bigcup_{x \in RD_c} \lambda(x)$ (Section~\ref{sec:kg_tier}) and the associativity, commutativity, and idempotence of set union.
\end{proof}

The proposition makes no use of how $S_{c'}$ arises; it holds for every set of cells.
The proposition, therefore, extends from the local to the propagated modules that roll-up delivers, by applying it to the coverage closure of the collapsing cells.

\begin{proposition}[Construction-time merge with propagation]\label{prop:merge-propagation}
Let $S^{\preceq}_{c'} = \{\, c'' \in F \mid c \preceq c'' \text{ for some } c \in S_{c'} \,\}$ be the coverage closure of $S_{c'}$. For every target cell $c'$,
\[
K'_{c'} \;=\; \bigcup_{c \,\in\, S_{c'}} K^{\preceq}_c
\;=\; \bigcup_{c'' \,\in\, S^{\preceq}_{c'}} K_{c''}
\;=\; \bigcup_{x \,\in\, RD_{S^{\preceq}_{c'}}} \lambda(x),
\]
that is, the rolled-up module, inherited knowledge included, equals one construction from the combined raw data items of the collapsing cells and of every cell covering them.
\end{proposition}

\begin{proof}
The second equality unfolds the propagated modules into local modules over the closure; the third applies Proposition~\ref{prop:merge} to $S^{\preceq}_{c'}$.
\end{proof}

Because $F_{\mathbf{e}}$ is closed under coverage, $S^{\preceq}_{c'}$ lies within the sliced subcube. Resolution already returns every cell the closure needs, and the construction path draws the raw data items of the covering cells alongside those of the covered ones.

Two consequences follow. First, the system may compute each source cell's final coordinates before any module exists and lift each raw data item exactly once into its merged target. No intermediate cube is built, merged, and discarded. Second, a merged module assembled from the cached modules of individual cells is identical to one constructed afresh, which is the composition half of the cache's correctness argument (Section~\ref{sec:kg_tier}).
Both identities concern asserted statements. When reasoning is requested, the derived knowledge is computed over the extended modules $M^{\preceq}_c = K^{\preceq}_c \cup \Tcal_\Sigma$ (Section~\ref{sec:kg_tier}) and kept separate from the asserted modules.

\subsubsection{Knowledge Graph Transformation}\label{sec:transformation}

The constructed cube is a set of pairs $\langle c, K_c \rangle$, a logical object that consuming applications cannot process directly.
The latter two flatten the modules into a single graph suited to whole-scope analytics.
The three transformations are invoked in the same way. Each consumes the same stream of constructed statements and differs only in what it emits.

\paragraph{RDF quads.}
The quad representation is the canonical one, since it loses nothing of the cube.
Each cell $c$ is identified by a context resource, and the assertions of $K_c$ occupy a named graph associated with that resource, its \emph{asserted module}.
Metadata statements record the structure. Each context declares its asserted module. When reasoning is requested, the derived assertions occupy a second, separate module, linked to the context and to the asserted module from which it was derived.
Coverage between contexts is likewise stated explicitly.
The vocabulary for this metadata is that of the CKR reference implementation~\cite{Serafini2012,Bozzato2013}, so the output can be loaded by tooling built for contextualized knowledge repositories and for KG-OLAP~\cite{schuetz2021}.

\paragraph{Labeled property graph.}
The system maps statements onto nodes, edges, and properties by three rules of the kind established for RDF to property graphs~\cite{angles2020mapping}: a typing statement sets the label of the subject's node; a statement with a literal object becomes a property of the subject's node; and a statement connecting two resources becomes an edge between their nodes. Labels, property keys, and edge labels are each the final segment of the corresponding identifier.
Nodes are identified by resource identifier. A node without a typing statement carries a generic label.
The named-graph tags are not carried over. The modules of the selected cells merge into one property graph, in which the context resources and their metadata appear as ordinary nodes and edges.

\paragraph{GraphFrame.}
The GraphFrame representation applies the rules of the labeled property graph, but emits the result as two tables, one of nodes and one of edges, serialized so that a Spark session can load them directly and reconstruct a GraphFrame.
This representation targets construction results too large for a single client, where the consumer is a compute cluster.

\subsection{Frontend Tier}\label{sec:frontend}

The \emph{Frontend Tier} receives KG specifications and delivers the constructed cubes.
The tier adds no knowledge processing of its own.
The OLAP Tier already delivers the cube in standard representations (Section~\ref{sec:olap}), so the frontend's task is access rather than translation. Any tool that reads those representations becomes, in effect, a frontend of the KG Lakehouse.

The Frontend tier exposes a single entry point, and that entry point serves both directions of use.
A client registers a cube schema and submits documents for ingestion. To query, it submits a KG specification and receives the constructed CKG as a stream, in the representation the specification names.
Interactive use is not a separate path. An interactive client holds no data and carries no logic of its own, and reaches the system through the same entry point as any program.
Nothing about that entry point is specific to the KG Lakehouse, so any programming environment can act as a client.

The frontend ends where standard representations begin. Established ecosystems carry the analytics from there.
The quad representation feeds standard Semantic Web tooling. For a pre-flight briefing, a specification names one route and one day. The returned cube loads into a triple store, and briefing rules query with SPARQL exactly the knowledge in scope, the runway closures and weather warnings along that route. Such rule-based briefing generation is established practice~\cite{Steiner2016,Schuetz2019}.
The property-graph representation feeds graph databases and their analytics. Examples are a controller-support view of one region during the validity of a significant-weather message, and conflict detection that joins traffic with contextual knowledge~\cite{vetter2024}.
The GraphFrame representation feeds cluster analytics. A specification wide enough for a season of flights yields the input for distributed feature extraction in trajectory prediction~\cite{Georgiou2020}.
Finally, the consumer need not be a person or a batch job at all. 
For example, adaptive ATM software can maintain its situational picture by continuously requesting small, current CKGs~\cite{ahmad2025ckglakehouse}. 
On-demand construction makes that pattern affordable, because the cost of each request depends on the scope of the query rather than the size of the corpus.

\section{Implementation}\label{sec:poc}
\begin{figure*}
    \centering
    \includegraphics[width=\textwidth]{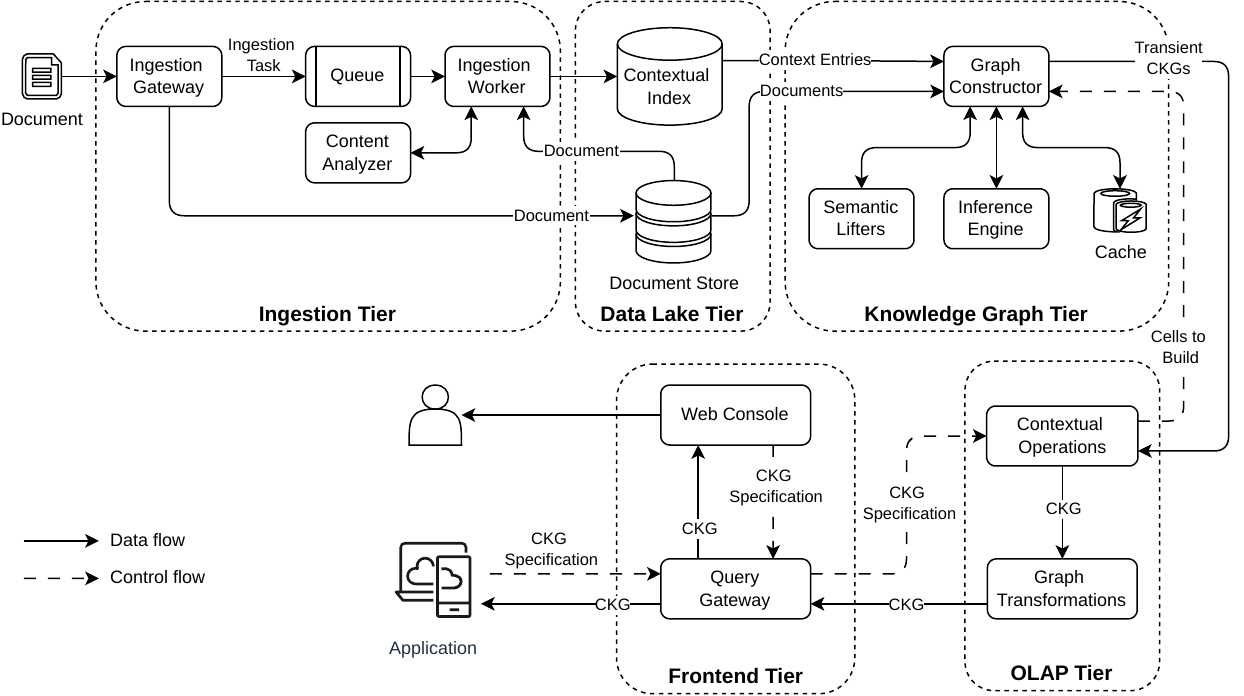}
    \caption{High-level architecture of the KG Lakehouse implementation}
    \label{fig:high_level_architecture}
\end{figure*}

The proof-of-concept implementation of the KG Lakehouse demonstrates that the architecture of Section~\ref{sec:architecture}, with its commitments to on-demand construction and to domain independence, is realizable with established cloud technology, and serves as the instrument of the evaluation in Section~\ref{sec:evaluation}.
Figure~\ref{fig:high_level_architecture} shows the components of the implementation and their interactions. 
The remainder of this section explains how each component is realized, and why.
The implementation consists of five independently deployable services, three format engines, and the libraries they share, written in Java and deployed as
container images on Kubernetes~\cite{K8s}.
The source code, the deployment configuration, and a set of small worked examples are openly available~\cite{kglakehouse}.

\paragraph{Services and communication.}
Five services divide the work: a client-facing gateway, the \emph{Ingestion Service}, the \emph{Query Service}, the \emph{Graph Service}, and the optional \emph{Inference Service}.
The gateway realizes the logical entry points of the architecture in one service, reached over HTTP. It receives documents on behalf of the Ingestion Tier (Section~\ref{sec:backend_tier}) and serves as the \emph{Query Gateway}, the entry point of the Frontend Tier (Section~\ref{sec:frontend}). The proof of concept also ships the \emph{Web Console}, an ordinary client of that entry point, which presents a returned cube as an interactive graph, a table, a cube view, or the raw serialization, and offers operational views of query history, performance, and service health. The services are stateless and share no database, so each scales horizontally on its own bottleneck. They communicate in two ways only.
Data ingestion is decoupled asynchronously using a message queue (Apache Kafka~\cite{Kafka,10010390}). The queue buffers bursts of arriving documents, while a pool of consumers processes the backlog. Indexing throughput therefore scales by adding consumers, without modifying other components. Conversely, query construction is coupled synchronously via remote procedure calls~\cite{gRPC}. Since queries are interactive and latency-sensitive, the query execution path streams results to the user as they are generated. The split reflects the different cost profiles of the two workloads.

\paragraph{Storage.} 
The storage architecture consists of three data stores, each selected for its access pattern and isolated behind a narrow interface.
First, an object store (MinIO~\cite{MinIO}) retains raw documents exactly as they arrive, serving as the authoritative copy of the raw data.
Second, the contextual index resides in a wide-column store (Apache Cassandra~\cite{Cassandra}). Its partitioned key-value model matches the index's access pattern. Selection needs a single lookup per dimension member followed by an intersection, without joins or scans. Its horizontal scalability accommodates the index's growth.
Third, an in-memory store (Redis~\cite{Redis}) holds constructed knowledge modules. It is the shared level of a two-level cache. In front of it, each \emph{Graph Service} replica keeps a bounded in-process cache, which spares a repeated request its network round trip, while the shared store lets every replica reuse a module another has constructed. Two safeguards bound what a fault in the cache can cost. Each entry expires after a fixed time, so an entry that an invalidation announcement failed to evict cannot outlive that bound, and a request the shared store cannot serve falls back to construction from the documents. Clients never query these stores directly. The stores are reached only through the system's services.

\begin{figure*}
    \centering
    \includegraphics[width=\textwidth]{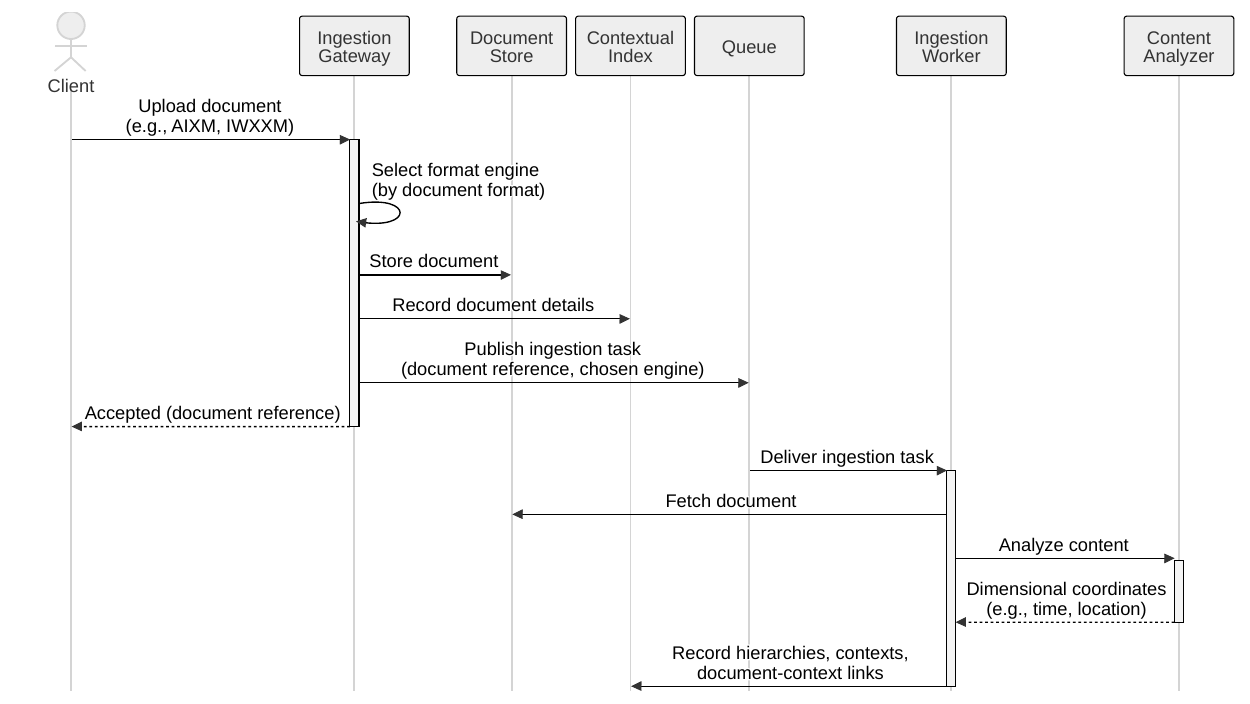}
    \caption{Sequence diagram of document ingestion and indexing}
    \label{fig:ingestion_flow}
\end{figure*}

\paragraph{Ingestion.}
Two steps take a document through the ingestion pipeline (Figure~\ref{fig:ingestion_flow}). First, the gateway accepts an uploaded document, persists it unchanged to the object store, and publishes an ingestion task to the message queue. Acceptance is thus decoupled from indexing, and the client request returns immediately.
Next, an ingestion worker consumes the task, retrieves the document, and passes it to the format-specific content analyzer. The analyzer answers a single question: which contexts does this document discuss? The worker then records this response in the index, capturing the member hierarchies of each dimension, the contexts as hashed coordinates, and the links mapping each context back to its source documents. 
The pipeline produces no graph statements at this stage. The pipeline's entire output is metadata.

\begin{figure*}
    \centering
    \includegraphics[width=\textwidth]{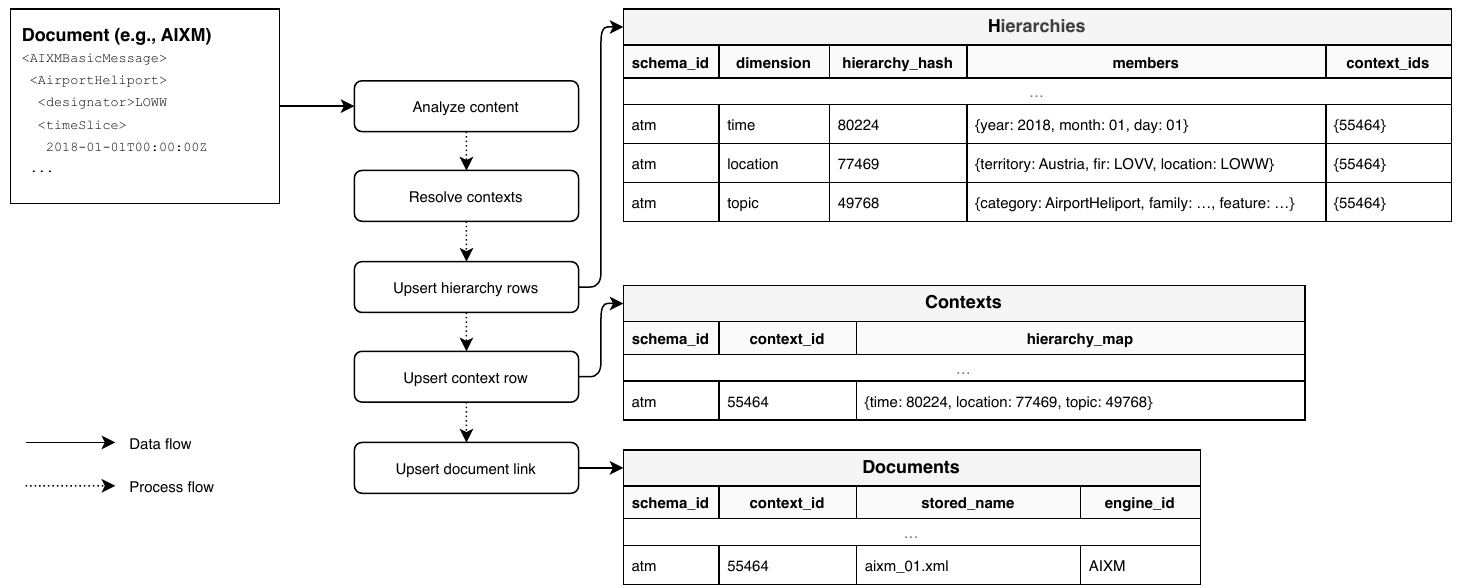}
    \caption{Contextual indexing by example}
    \label{fig:index_example}
\end{figure*}

\paragraph{Contextual index.}
Three tables carry the contextual index: \emph{hierarchies}, \emph{contexts}, and \emph{documents}.
Figure~\ref{fig:index_example} traces an example document (\texttt{aixm\_01.xml}) through those tables. 
Per dimension, the \emph{hierarchies} table records the member chains that occur in the data. 
A member chain is a root-to-leaf path through the dimension's hierarchy, e.g., the year 2018, the first month of year, and the first day of month in the time dimension, or the Austrian territory, the LOVV flight information region, and the LOWW aerdrome in the location dimension. 
The identifier of a member chain is its hash; the hash is the key of the hierarchies table.
For each member chain, the hierarchies table stores the set of contexts where the chain occurs.
The hierarchies table is partitioned by dimension, so resolving one dimension of a query scope costs a single partition lookup.
The \emph{contexts} table names the cells.
A cell's identifier is derived from the hash of its coordinates, the combination of one chain per dimension, so two documents that concern the same coordinates meet in the same cell without coordination among ingestion workers. 
The \emph{documents} table links each cell to the stored documents behind it, together with the format engine that lifts them. 
The documents table is partitioned by cell, so collecting a cell's documents is again a single lookup. 
These three lookups---of member, cell, and documents---are the entire read path of selection. The index resolves a scope without a join or a scan.

Context assignment is computed at the granularity of individual data items, since a single document can span multiple contexts, but the index records are linked at the granularity of documents. 
This asymmetry shapes both the index and the construction path. 
Consequently, the index grows with the number of documents---context links rather than with the volume of document content---and remains a small fraction of the size of the corpus (Section~\ref{sec:evaluation}). 
This space efficiency comes at a read-time cost. 
The construction path must parse every document that is loaded a second time, to recover the document's data items. 
The trade-off is nonetheless favorable. 
Content extraction runs lazily, only for the cells a query targets, while the index, the only structure that must span the entire corpus, stays compact.

\begin{figure*}
    \centering
    \includegraphics[width=\textwidth]{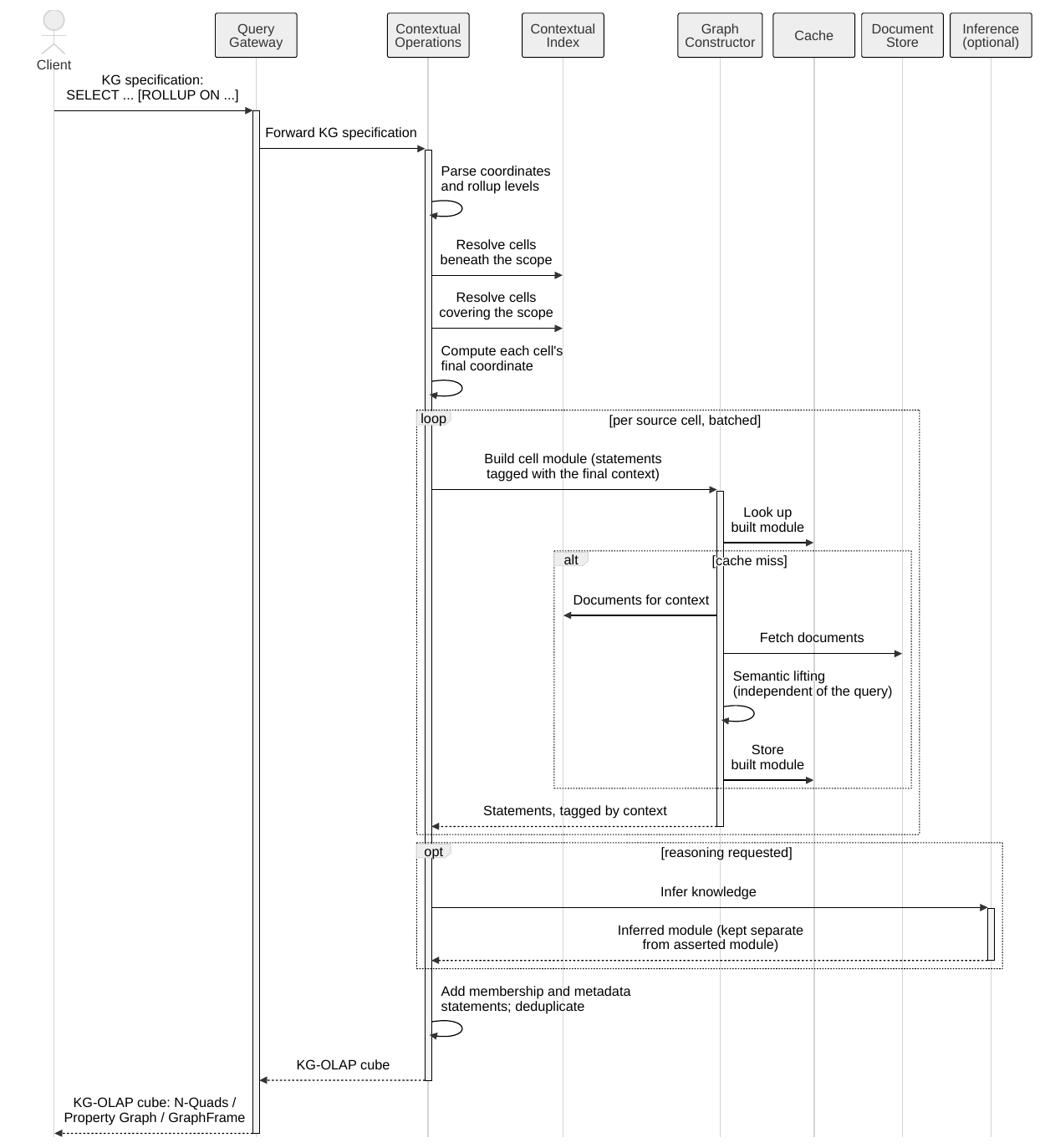}
    \caption{Sequence diagram of on-demand knowledge graph construction}
    \label{fig:construction_flow}
\end{figure*}

\paragraph{Construction.}
A KG specification drives construction in two stages (Figure~\ref{fig:construction_flow}): planning by the \emph{Query Service} and module construction by the \emph{Graph Service}. The \emph{Query Service} parses the specification into one coordinate per dimension and, optionally, a target roll-up level per dimension.
The Query Service resolves the scope against the index in both directions: the cells at or beneath the requested coordinates, and the coarser cells covering them, whose knowledge propagates downward (Section~\ref{sec:olap}). 
The Query Service then computes each cell's final coordinates under the roll-up. 
A cell's identity is the hash of its coordinates, so cells that roll up to the same coordinates collapse into one.
The merge needs no bookkeeping beyond the hash. 
The service issues one construction request per source cell, which can be processed in parallel.

For each source cell with a construction request, the \emph{Graph Service} loads the cell's documents, runs the format's semantic lifter, and streams the resulting statements back, tagged with the cell's final context. 
Repeated requests are answered from the cache. 
Lifting depends only on the document, which the query never reaches, so construction is deterministic, and the cache can only save time, never change a result. 
The streamed statements pass through a duplicate filter, keyed by a hash of each statement's content, which makes the answer a set, independent of the order in which cells complete.
A specification selects one of three output models, produced by interchangeable graph builders behind a single lifting step. 
RDF quads~\cite{RDF} feed triple stores such as Fuseki~\cite{fuseki}.
A labeled property graph feeds graph databases such as Neo4j~\cite{neo4j}. 
A GraphFrame~\cite{graphframes} serialization feeds Spark for distributed processing.

\begin{figure}
    \centering
    \includegraphics[width=\columnwidth]{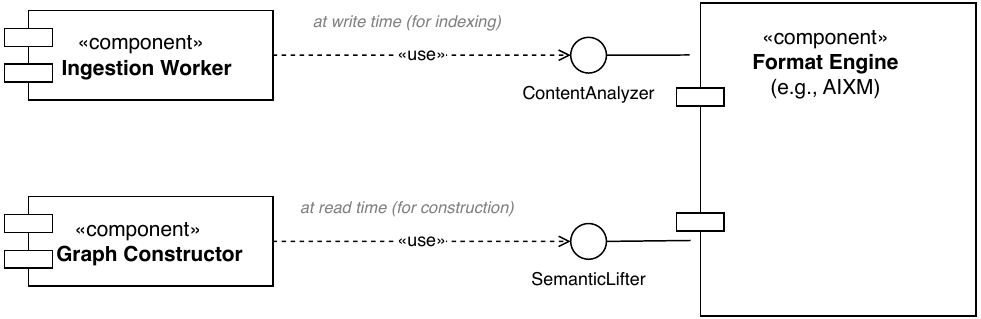}
    \caption{Component diagram of the format engine}
    \label{fig:format_engine}
\end{figure}

\paragraph{Format engines.}
All format knowledge lives in engines, one per source format, each contributing exactly two functions: a content analyzer used at write time and a semantic lifter used at read time (Figure~\ref{fig:format_engine}). 
The proof of concept comes with three engines: for AIXM~\cite{AIXM2019}, IWXXM~\cite{IWXXM2019},
and FIXM~\cite{FIXM2019}.
No service has the engines hard-wired in the source code; the engines are discovered at run time. 
Supporting a further format is the work of writing an analyzer and a lifter for the format.
Consequently, moving to another domain is a change of configuration, plus engines for its formats where none exist.
No assumptions from the aeronautical domain are compiled into the services or the stores. 
The cube schema, with its dimensions, levels, and member hierarchies, is declared as configuration and registered by a client (Section~\ref{sec:frontend}), so the aeronautical character of the proof of concept rests on that declaration and on its three engines alone.
The system's understanding of a format is therefore applied when a document is read, not when the document is stored, which is the schema-on-read nature of the lakehouse~\cite{Armbrust2021}.

\paragraph{Reasoning.}
Ontological reasoning is optional and isolated in the \emph{Inference Service}.
When a KG specification explicitly requests reasoning, the constructed module of each cell is passed to a rule engine (Apache Jena~\cite{jenaFramework}), together with the terminological knowledge, to infer additional statements.
These derived statements are returned in a separate module, so they never mix with the asserted facts.
The terminology of the proof of concept is deliberately modest, a subclass hierarchy derived from the levels of the cube schema, and the reasoner is a standard RDFS or OWL profile. Because the service receives a constructed module and returns derived statements, a richer ontology or a more capable reasoner can take their place without changing the construction path.
Factoring reasoning out of the construction path keeps that path free of reasoning's overhead and allows this computationally demanding step to scale independently.

\paragraph{Observability.}
The implementation doubles as the instrument of the evaluation. Every stage publishes named measurements: scope-resolution time, per-cell construction time, merge time, cache hits and misses, output size. A request also carries an identifier across service boundaries, so a single query can be followed end to end.
Section~\ref{sec:evaluation} reports exactly these measurements, read from the running system.

\paragraph{Model and realization.}
Each construct of the formal model has a concrete counterpart in the implementation. 
Table~\ref{tab:model_realization} pairs each construct from Section~\ref{sec:kgolap} and the architectural tiers with the realizing mechanism in the proof-of-concept implementation.

\begin{table*}
\caption{The formal model and its realization in the implementation}
\label{tab:model_realization}
\begin{tabularx}{\textwidth}{lX}
\toprule
\textbf{Formal construct} & \textbf{Realization} \\
\midrule
Cube schema $\Omega$ & declarative configuration of dimensions; member hierarchies recorded in index \\
Extraction $\phi$ and context assignment $\kappa$ & dedicated content analyzer per format called at write time \\
Cell name $\mathsf{cn}(\mathbf{d})$ & hash of the dimensional coordinates \\
Contextual index $\Idx$ & \emph{hierarchies}, \emph{contexts}, and \emph{documents} tables in the \emph{Index Store} \\
Coverage $\preceq$ & ancestry check over the member chains recorded in the index \\
Retrieval $RD_c$ & index lookup, then the data items are extracted again from each document \\
Semantic lifting $\lambda$ & dedicated semantic lifter per format called at read time \\
Local module $K_c$ & statements produced for one cell in the requested representation \\
Roll-up (merge) & coordinates coarsened before construction, so that equal cells merge \\
Reasoning over a module & optional \emph{Inference Service}, returning a separate derived module per cell \\
\bottomrule
\end{tabularx}
\end{table*}

\section{Evaluation}\label{sec:evaluation}
The evaluation asks whether the implemented artifact exhibits the properties the architecture was designed to deliver, under controlled and instrumented conditions.
The evaluation comprises five experiments, 
E1--E5, each answering one question (Table~\ref{tab:experiments}), 
and stresses the system along the Three Vs of big data: volume, velocity, and variety~\cite{Laney2001}.
This section reports the design and the principal results. 
The full protocols, per-axis tables, and figures, together with the deployment manifests, scripts, and measured data, are in the online repository~\cite{kglakehouseeval}.

\begin{table*}
\caption{The five experiments, the questions each answers, and the part of the system each evaluates}
\label{tab:experiments}
\begin{tabularx}{\textwidth}{cXl}
\toprule
 & \textbf{Questions} & \textbf{Component evaluated} \\
\midrule
\multirow{4}{*}{E1} & 
  \begin{itemize}[nosep,leftmargin=*]
    \item Does the ingestion pipeline keep pace with heterogeneous, high-velocity data at scale?
    
    \item How soon are ingested documents available to queries? 
  \end{itemize} & 
\multirow{4}{*}{Ingestion pipeline} \\

\multirow{4}{*}{E2} & 
  \begin{itemize}[nosep,leftmargin=*]
    \item Does the cost of context selection follow query scope rather than corpus size? 

    \item Is the index compact?
  \end{itemize} & 
\multirow{4}{*}{Contextual index} \\

\multirow{3}{*}{E3} & 
  \begin{itemize}[nosep,leftmargin=*]
    \item Does the cost of on-demand construction follow query scope rather than corpus size? 
  \end{itemize} & 
  \multirow{3}{*}{Construction path} \\
  
\multirow{4}{*}{E4} & 
  \begin{itemize}[nosep,leftmargin=*]
    \item Are the implemented contextual operations and graph transformations semantically correct over reconstructed graphs (at every hierarchy depth)? 
  \end{itemize} & 
\multirow{4}{*}{Contextual operations and graph transformations} \\

\multirow{4}{*}{E5} &
  \begin{itemize}[nosep,leftmargin=*]
    \item What does each approach cost, in terms of run time and memory consumption, to return the knowledge modules for a queried context?

    \item What limits the scale of each approach?
  \end{itemize} &
\multirow{4}{*}{Whole system vs.\ materialized baseline} \\
\bottomrule
\end{tabularx}
\end{table*}

\subsection{Method and Datasets}\label{sec:datasets}

Correctness is established before performance is measured, and by independent means. Whether an operation returns the right result is decided by an \emph{oracle}.
The operands are loaded into a separate triple store, an Apache Jena TDB2 dataset served through Fuseki~\cite{fuseki}, and compared by the emptiness of their symmetric difference. The store plays no part in constructing the answers, so a fault in our own implementation cannot mask a mismatch.
Performance is then measured on a distributed deployment of the system (the five services of Section~\ref{sec:poc} on a multi-node Kubernetes cluster), with every stage instrumented. The quantities reported below are read from the running system.

All experiments draw on one deterministic synthetic dataset family, produced by the project's data generator from a fixed, seeded aviation scenario~\cite{atmgen}, so every dataset is exactly reproducible and the correct context of every statement is known by construction. 
The datasets follow a one-factor-at-a-time design~\cite{montgomery2019}. Each moves a single axis off a fixed center (a country-scale geography, a thirty-day horizon, one thousand flights per simulated day, all three source formats, and all three dimensional hierarchies), holding the others fixed, so that an observed effect is attributable to one factor. The axes are geographic scope (one aerodrome to a continent), operational density (one hundred to ten thousand flights per day), and temporal horizon (one day to a year). The corpus spans roughly an order of magnitude in document count within one operational scenario. Variety is supplied by the three exchange standards of the use case (AIXM, IWXXM, and FIXM~\cite{AIXM2019,IWXXM2019,FIXM2019}), ingested and queried through the same plugin mechanism.

\subsection{Results}

Together, the five experiments establish that on-demand construction returns the graphs the model prescribes, at a cost set by the scope of the query rather than the size of the corpus.
Each experiment rests on the result before it. 
E1 shows that ingestion and indexing keep pace with arriving data, so that a query always reads a current index. 
E2 finds that the first stage of answering a query, resolving it to the contexts it names, costs according to the scope of the query rather than the size of the corpus, and that the index supporting it remains a small fraction of that corpus.
E3 carries the same test to construction, which dominates the cost of a query, and finds construction unmoved by a growing corpus but responsive to a widening scope. 
E4 confirms that the graphs so constructed are correct at every hierarchy depth. 
E5 sets the two approaches side by side and measures what each costs to return the same knowledge modules, and what limits the scale of each.

\begin{figure}
    \centering
    \includegraphics[width=\columnwidth]{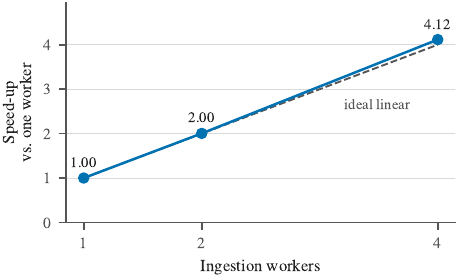}
    \caption{E1, capacity axis: the indexing lag halves as the ingestion workers double.}
    \label{fig:eval_e1_capacity}
\end{figure}

\paragraph{E1: Ingestion and indexing.}
We examine whether the ingestion pipeline keeps pace with heterogeneous, high-velocity data at scale and how soon ingested documents become available to queries.
The pipeline pays a constant per-document cost.
As operational density rises a hundredfold, growing the corpus from \num{28590} to \num{324456} documents,
the pipeline processes each document in an essentially constant 27--28\,ms.
Accumulating the three formats leaves that cost unchanged, so heterogeneity carries no per-document penalty.
Throughput reflects the offered load rather than a ceiling, because the client drives ingestion at a fixed rate. Capacity is therefore read from the \emph{indexing lag}, the interval from the acceptance of the last document of an offered batch until the whole batch is indexed and available to queries. The lag is the time the pipeline needs to drain its backlog, and it falls in proportion as workers are added: on the center corpus of \num{55580} documents, from \num{1607} seconds with one worker to 802 with two and 390 with four, a speed-up of 4.12 that is essentially linear (Figure~\ref{fig:eval_e1_capacity}). Notices carry a cost of their own, and it falls on the size of the index rather than on ingestion speed. A notice to airmen is valid over a span of days, about six on average, and the pipeline records one context for every day it covers, so 156 such documents expand to roughly 941 index entries.

\begin{figure}
    \centering
    \includegraphics[width=\columnwidth]{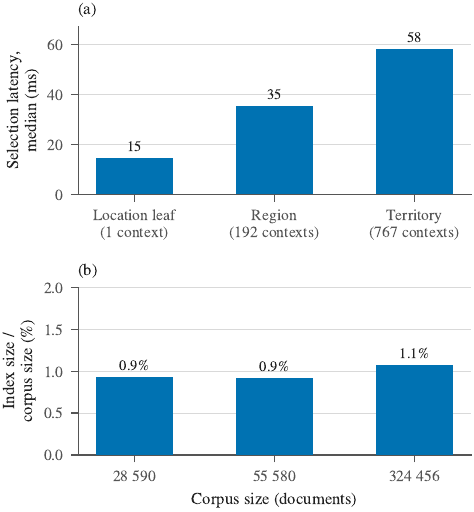}
    \caption{E2: (a)~selection latency (median of three repeats) rises about fourfold while the
    matched context set widens 767-fold; (b)~the index stays near one percent of the corpus as the
    document count grows elevenfold. The whole-space query duplicates the territory scope and is omitted.}
    \label{fig:eval_e2}
\end{figure}

\paragraph{E2: Contextual indexing.}
We measure the cost of context selection and the size of the index that supports it. Selection is only the first stage of answering a query, the index lookup that turns a declared scope into the set of matching contexts, before any graph is built. Constructing those contexts' graphs is the subject of E3. The cost of resolving a query to its contexts follows the scope of the query, not the size of the corpus. As the matched context set widens from a single context to all 767, selection latency rises about fourfold, from 14.7 to 58\,ms (Figure~\ref{fig:eval_e2}a), and for a fixed scope it does not rise at all as the document count grows elevenfold. The index is compact. It runs from 2.4 to 19.5\,MB against a corpus of 263\,MB to 1.81\,GB, about one percent throughout (Figure~\ref{fig:eval_e2}b). Its structure (767 contexts) is invariant to volume. A query can therefore locate its slice quickly while the bulk of the data rests, untransformed, in low-cost storage.

\begin{figure}
    \centering
    \includegraphics[width=\columnwidth]{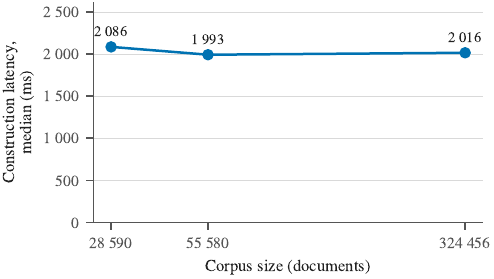}
    \caption{E3, volume axis: at a fixed query scope, construction latency (median of three
    repeats) is flat as the document count grows elevenfold.}
    \label{fig:eval_e3_volume}
\end{figure}

\begin{figure}
    \centering
    \includegraphics[width=\columnwidth]{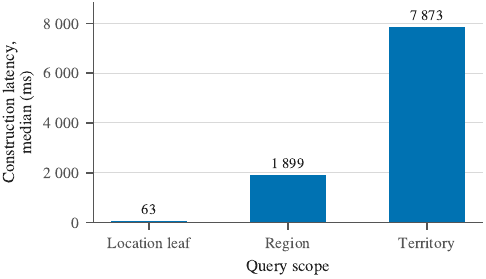}
    \caption{E3, selectivity axis: at a fixed corpus, construction latency (median of three
    repeats) grows with the requested scope. The whole-space query is omitted, since it resolves to the
    same 767 contexts as the territory scope and costs the same.}
    \label{fig:eval_e3_selectivity}
\end{figure}

\paragraph{E3: On-demand construction.}
We investigate whether the cost of constructing a requested CKG follows the scope of the query or the size of the corpus, varying each in turn. At a fixed query scope, one region, construction builds the same graph (188 contexts, \num{56474} quads) and takes essentially the same time (\num{2086}, \num{1993}, and \num{2016}\,ms; a largest-to-smallest ratio of 1.05) as the surrounding corpus grows from \num{28590} to \num{324456} documents (Figure~\ref{fig:eval_e3_volume}).
The identical quad count across corpus sizes shows the mechanism. The same scope resolves to the same contexts and does the same work, and the eleven-times larger corpus contributes nothing to the cost. Conversely, at a fixed corpus, construction grows with the requested scope, from 63\,ms for a single-leaf query of 425 quads to 7.9\,s for a territory-wide query of \num{227039} quads (Figure~\ref{fig:eval_e3_selectivity}). Cost depends on the scope of the query, not the size of the corpus.
The marginal cost per context is steady. Beyond the smallest scope, construction takes 10--11\,ms per context (188 contexts in about 2\,s; 767 contexts in 7.9\,s) at a throughput near \num{28000} quads per second on both axes, while the 63\,ms single-leaf query reflects a fixed per-query floor.

\paragraph{E4: Correctness of the KG-OLAP operations.}
We verify the contextual operations and graph transformations over reconstructed graphs, at every hierarchy depth. Correctness is checked through properties that must hold between two of the system's own answers (slice containment, roll-up as the union of covered cells, roll-up transitivity, dice commutativity, drill-down as inverse roll-up, and covering delivery), each decided by the independent oracle on exact set comparison, with reasoning off, so that the checks isolate the operations themselves from inference.
The covering-delivery property checks knowledge propagation. The knowledge of a context that covers cells inside the queried scope (coarser than the scope in one dimension and finer in another) must reach the answer. Because the corpus lies within one calendar month, constraining any slice by that month excludes nothing, so the constrained and unconstrained answers must coincide. A resolution that dropped such covering contexts would fail the equality.
Fourteen checks apply across the three hierarchy depths (six at depth three, five at depth two, three at depth one), and all fourteen pass (Table~\ref{tab:e4}). The number of applicable properties falls as the hierarchy flattens, exactly as the model requires: transitivity needs three levels, roll-up and drill-down two, while covering delivery applies at every depth. That is itself a check on the operations. The graphs that E3 builds cheaply are therefore also correct.

\begin{table}
    \caption{E4: All fourteen correctness checks pass against the independent oracle. A dash marks a property that cannot exist at that depth.}
\label{tab:e4}
\centering
\begin{tabularx}{\columnwidth}{Xccc}
\toprule
 & \multicolumn{3}{c}{\textbf{Hierarchy depth}} \\
\cmidrule(lr){2-4}
\textbf{Property} & \textbf{3} & \textbf{2} & \textbf{1} \\
\midrule
Slice containment & \checkmark & \checkmark & \checkmark \\
Roll-up as union of covered cells & \checkmark & \checkmark &---\\
Roll-up transitivity & \checkmark &---&---\\
Dice commutativity & \checkmark & \checkmark & \checkmark \\
Drill-down as inverse roll-up & \checkmark & \checkmark &---\\
Covering delivery & \checkmark & \checkmark & \checkmark \\
\bottomrule
\end{tabularx}
\end{table}

\paragraph{E5: On-demand construction versus monolithic materialization.}
We compare what the two approaches spend to serve the same task, returning the knowledge modules for a queried context. The baseline mirrors conventional practice: the corpus is mapped to RDF with a declarative mapping language (RML~\cite{dimou2014rml}), transformed into the full contextualized cube by a SPARQL query, and bulk-loaded into Apache Jena's TDB2 store~\cite{jenaFramework}. Both sides serve queries on identical commodity virtual machines. Table~\ref{tab:e5} states the comparison on the larger of E5's two corpora, 256 aerodromes over sixty days. The thirty-day corpus supplies the scaling comparison below.
To answer any query at all, the baseline first builds the whole cube of 7.26 million quads, peaking at 46\,GB of memory over about fifty-five minutes. That peak exceeds the test machines, so the build was measured on a larger host. On-demand construction builds only the queried context, in tens of milliseconds per context and within about 1.4\,GB.
The costs differ in kind: one is a function of the whole knowledge graph, the other of the queried
scope.

\begin{table}
\caption{E5: the cost of first returning the knowledge modules for a queried context (256 aerodromes, 60 days). The materialized column includes the one-time cube build. Full tables in the
appendix~\cite{kglakehouseeval}.}
\label{tab:e5}
\begin{tabularx}{\columnwidth}{lXX}
\toprule
 & \textbf{Materialized} & \textbf{On-demand} \\
\midrule
What is built? & the whole cube (7.26\,M quads) & the queried context \\
Peak memory & 46\,GB & $\approx$~1.4\,GB \\
Time & $\approx$~55\,min & tens of ms per context \\
Grows with & the whole graph & the queried scope \\
\bottomrule
\end{tabularx}
\end{table}

\begin{figure}
    \centering
    \includegraphics[width=\columnwidth]{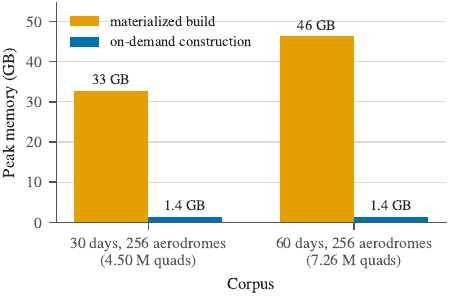}
    \caption{E5: the peak memory of the materialization build rises with the knowledge graph, while
    on-demand construction stays near 1.4\,GB.}
    \label{fig:eval_e5_wall}
\end{figure}

\begin{figure}
    \centering
    \includegraphics[width=\columnwidth]{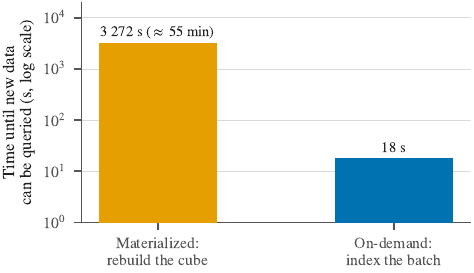}
    \caption{E5: rebuilding the materialized cube against indexing on demand, measured as the time until
    queries can answer from a batch of 255 newly arrived documents (logarithmic scale)}
    \label{fig:eval_e5_update}
\end{figure}

Three consequences follow from the difference in kind between the two costs.
The first consequence is feasibility. The memory of the whole-graph transformation rises with the graph, from 33 to 46\,GB as the corpus doubles from thirty to sixty days (Figure~\ref{fig:eval_e5_wall}), so the cube eventually cannot be built on any single machine, whereas on-demand construction has no whole-graph ceiling.
The second consequence is wasted work. The materialized cube holds \num{40511} contexts, of which a region-scoped query reads about four percent and an aerodrome-scoped query a single context. Materialization pays to build all of them.
The third consequence is the lag before new data can be queried. Indexing makes a batch of 255 newly arrived documents available to queries in about eighteen seconds, against the fifty-five-minute rebuild the materialized cube requires to reflect the same change (Figure~\ref{fig:eval_e5_update}).

\subsection{Discussion}

On-demand construction is a way to avoid maintaining a monolithic knowledge graph, not a more efficient way to build one. 
Some applications may genuinely require the entire graph to be materialized, including certain types of analytical questions, arbitrary cross-cutting queries, or a general-purpose SPARQL endpoint.
For such applications, a materialized triple store remains the appropriate solution. 
Once the graph has been materialized, answering a query from it is a lookup, whereas on-demand construction must still build the requested modules. For that task the architecture offers no advantage. 
Instead, the proposed architecture's contribution is orthogonal: The KG Lakehouse parallelizes scope-bounded construction, so that a query pays only for the knowledge modules of the contexts it names. 
Consequently, the architecture is intentionally optimized for on-demand construction rather than for materializing an entire KG.

Materialization does not scale without bound.
The hazards of maintaining a single, monolithic KG---namely the recognized cost of materializing an RDF graph, the need to arrange any partitioning before the data is loaded, and the maintenance risk of a store left effectively unmaintained---are laid out in Section~\ref{sec:related}.
A materialized store is nonetheless simpler to operate than an on-demand pipeline.
The advantage of on-demand construction increases as the corpus grows, its formats diversify, and its content changes, which are the conditions of the use case described in Section~\ref{sec:case}.

\section{Related Work}\label{sec:related}

A KG integrates heterogeneous information with ontological knowledge, which makes them useful across information retrieval, question answering, social network analysis, deep learning, and analytics~\cite{zou2016survay,Ali2021}.
The systems that manage them, however, are built around a standing graph, one that is assembled, loaded, and then queried.
In the following, we review related work on KG management systems, on cloud-native and lakehouse architectures, on graph storage and analytics, and on hybrid approaches, positioning the KG Lakehouse alongside each.

\subsection{Knowledge Graph Management Systems}
A KGMS holds a graph and answers queries over it. Neo4j~\cite{neo4j} does so for property graphs and Blazegraph~\cite{blazegraph} for RDF, offering traversal, pattern matching, and, in the RDF case, ontological inference.
Where one machine no longer suffices, distributed engines such as Apache Spark GraphX~\cite{graphx} and managed services such as Amazon Neptune~\cite{amazonNeptune} spread the work over several nodes, combining partitioning with distributed query execution and, in places, inference~\cite{Hogan2021}.
These systems share one assumption. The graph exists before a query arrives, and the query reads what the store already holds. None of them derives a graph in response to the query that needs it, which is the capability the KG Lakehouse provides. A further limitation is one of scale. A materialized graph must be built and held in full before it can be queried, and the cost of that construction, in time, processor, and memory, is a recognized concern in its own right, with dedicated benchmarks now measuring it across construction tools~\cite{krown}. Some stores do partition a graph across nodes~\cite{Ali2021,wylot2018rdf}, but the division must be arranged before the data is loaded, and queries that cross partitions pay for the split.
The Wikidata Query Service, the most prominent public knowledge graph deployment, ran on Blazegraph~\cite{blazegraph} as one unpartitioned graph of over sixteen billion statements. Reloading it from the data dumps takes one to two months, and it grows unstable near a size ceiling. In 2025 its operators divided it into separate endpoints, each queried on its own, to keep it operable~\cite{wdqs-split}, a difficulty compounded by Blazegraph itself having been effectively unmaintained since its developers were absorbed into Amazon Neptune~\cite{amazonNeptune}. Knowledge graph management systems remain well suited to serving a standing graph. They are not designed to avoid building the whole of it. The proposed KG Lakehouse avoids building the whole graph, not by materializing the monolith more efficiently (a task for which a purpose-built store remains appropriate) but by constructing only the contextualized subgraph a query requires.

\subsection{Cloud-Native and Lakehouse Architectures}
Cloud-native design, with containers and independently deployable services, lets the parts of a system scale separately, and the lakehouse paradigm~\cite{Armbrust2021} keeps data in its native format until a query reads it (Section~\ref{sec:intro}). Cloud-native design and the lakehouse paradigm combine naturally. Ingestion, indexing, and querying place different demands on a system, and each can be scaled on its own.
The Databricks Lakehouse Platform~\cite{Databricks} exemplifies this combination, integrating metadata management, caching, and indexing to support analytics across clouds, and Kona et al.~\cite{Kona2022} add semantic reasoning by combining Databricks with Stardog~\cite{Stardog}. Both work over tabular data. The KG Lakehouse carries the lakehouse paradigm to heterogeneous documents, from which it constructs contextualized KGs on demand.

\subsection{Advancements in KG Storage and Analytics}
Two lines of work bring graphs closer to the data lake. GraphAr~\cite{graphar2023} defines a storage layout for labeled property graphs on Parquet files, so that neighbor retrieval and label filtering run directly over lake storage. Its concern is how a graph is laid out at rest, whereas the KG Lakehouse leaves the data unlifted until a query names the contexts it needs. Stream2Graph~\cite{stream2graph2023} builds graphs continuously from streaming data and updates them online. Both maintain a standing graph. Neither derives, per query, the contextual subgraph that an analytical task requires.
Work on extraction is complementary rather than competing. Chen et al.~\cite{chen2020} build knowledge graphs from unstructured text with deep active learning, which lowers the amount of labeled data required. A method of this kind could act as one more content analyzer in the ingestion pipeline, where extraction is a plug-in step.
Several frameworks specifically address large-scale graph processing and analytics. Among the first was Graph OLAP~\cite{Chen2009}, which associates graphs with dimensional attributes, yielding a cube whose cells contain graphs, and distinguishes informational roll-up, which overlays snapshots of a network, from topological roll-up, which merges nodes within one. Graph Cube~\cite{Zhao2011} carries this view to warehousing multidimensional networks. In these frameworks, dimensional structure comes from the attributes of snapshots, nodes, and edges, the measures are aggregated graphs and numeric summaries over them, and the analyzed graph is presumed materialized. Contexts play no part. KG-OLAP~\cite{schuetz2021} instead attaches the dimensions to the knowledge itself, and the proposed method follows this route. KG-OLAP, however, operates on a materialized cube, whose cells hold knowledge modules lifted and stored before any query arrives. The KG Lakehouse leaves the cube mapping unevaluated until a query names cells, so that roll-up aggregates the knowledge modules of contexts over graphs constructed on demand rather than held in a warehouse. P-OLAP~\cite{Beheshti2015} applies graph-OLAP principles to process execution data, scaling the computation with Hadoop and MapReduce~\cite{Hadop}.
Large-scale RDF processing has a stack of its own: SANSA~\cite{Janev2020} layers SPARQL evaluation, reasoning, and machine learning on Apache Spark, and Trinity~\cite{Shao2013} serves online and offline queries from distributed memory. Both operate on a dataset that has already been assembled and loaded, which is the step the KG Lakehouse defers to query time. SANSA and Trinity are nonetheless compatible with the KG Lakehouse. A constructed cube can be handed to either for downstream analysis, since the GraphFrame representation of Section~\ref{sec:olap} targets the same Spark ecosystem.

\subsection{Hybrid and Domain-Specific Approaches}
A final group combines graph reasoning with big-data infrastructure. Wikibase~\cite{WikimediaDeutschland} offers open-source infrastructure for managing pre-existing KGs but lacks support for generating task-specific contextualized KGs on demand.
Federated SPARQL~\cite{endris2020federated}, linked data fragments~\cite{Verborgh2014}, and virtual KGs~\cite{Xiao2019} also avoid holding one graph in one place, but they answer queries by delegating them to remote endpoints, to fragment servers, or to the relational sources behind a mapping. The KG Lakehouse instead builds the requested subgraph itself, from raw documents it has indexed by context, and so depends on neither pre-existing triple stores nor query federation.
Gassauer-Fleissner et al.~\cite{GassauerFleissner2022} pair a cloud-native platform with a graph database to detect financial crime, an example of the same pairing in a domain where the relevant subgraph is likewise a small part of the whole.
Across these lines of work, the graph is stored, updated, or federated, but it is always there to be queried. The KG Lakehouse takes the opposite starting point. It keeps documents as they arrive, records the contexts in which their content is valid, and derives a knowledge graph only when a query names one. What that costs, and how it compares with building the graph in advance, is the subject of Section~\ref{sec:evaluation}.

\section{Conclusion}\label{sec:conc}

This paper presented the Knowledge Graph Lakehouse, a cloud-native architecture that carries the lakehouse paradigm to the management of contextualized knowledge graphs.
It is built for workloads whose scope is declared before any data is touched, for example a time window, a region, or a topic.
For such workloads the system inverts the usual order of knowledge-graph management. Rather than materializing a standing graph and querying it, it stores raw documents as they arrive, indexes the contexts in which their content is valid, and constructs on demand exactly the contextualized knowledge graph a query names, never the whole of it.

The architecture assigns the work to five tiers, grounded throughout in the KG-OLAP formalism.
That grounding has practical consequences. Because a module is determined solely by the indexed raw data, construction is deterministic and merging commutes with construction. Caching therefore becomes a partial materialization of the cube mapping that can save time without changing a result.

The architecture is realized as a proof of concept: five independently deployable services over an object store that keeps the raw documents, a wide-column store that holds the contextual index, a message queue that absorbs bursts of arriving documents, and an in-memory store the services share as a cache, all operated on a container platform.
The realization keeps the domain out of the software. The cube schema, with its dimensions, levels, and member hierarchies, is declared as configuration, and every source format is served by its own content analyzer and semantic lifter, so the aeronautical character of the proof of concept rests on that declaration and those engines alone.
The implementation is also the instrument of the evaluation, and it is openly available~\cite{kglakehouse}.

The evaluation supports the central proposition that construction cost depends on the scope of the query, not the size of the corpus.
As the corpus grew from \num{28590} to \num{324456} documents, the ingestion pipeline processed each document at an essentially constant cost. At a fixed query scope, construction returned the same graph in essentially the same time, with a largest-to-smallest ratio of 1.05 across an elevenfold corpus range. At a fixed corpus, cost rose with scope, from tens of milliseconds for a single-leaf query to seconds for a territory-wide query.
A cube materialized in advance holds tens of thousands of contexts, of which a scoped query reads a few percent or less. That is work the KG Lakehouse never performs.
Correctness was established independently of performance, by comparing reconstructed graphs against an independent oracle, a separate triple store, at every hierarchy depth.

The KG Lakehouse suits workloads whose scope is declared in advance, where it builds no more of the graph than a query names.
Arbitrary queries over a standing graph are better served by a materialized store.
Several directions remain open.
The lake is append-only. A correction arrives as a new document, and the assertions of both versions coexist in the affected modules. This is adequate for temporally versioned aeronautical data, where validity is part of the content. Retraction of a document, and of the knowledge lifted from it, remains future work.
The formal treatment of the contextual operations is to be completed with the interplay of merging, propagation, and reasoning.
Finally, the schema-on-read discipline leaves the system open to new data. Supporting a further source format, or another domain altogether, is the work of writing a content analyzer and a semantic lifter. Connecting the delivered representations to distributed knowledge-graph processing stacks is a natural continuation.

\section*{Acknowledgments}
This work was conducted as part of the AWARE project, which has received funding from the SESAR Joint Undertaking under grant agreement no. 101167442 within the European Union's Horizon Europe research and innovation program.
The views expressed in this work are those of the authors.

\begin{center}
  \includegraphics[width=.27\textwidth]{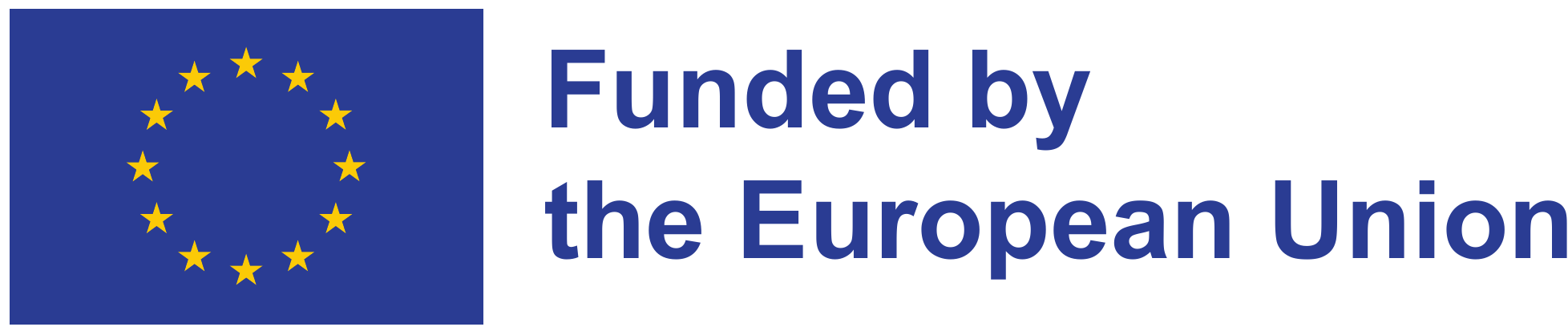}
  \hfill
  \includegraphics[width=.135\textwidth]{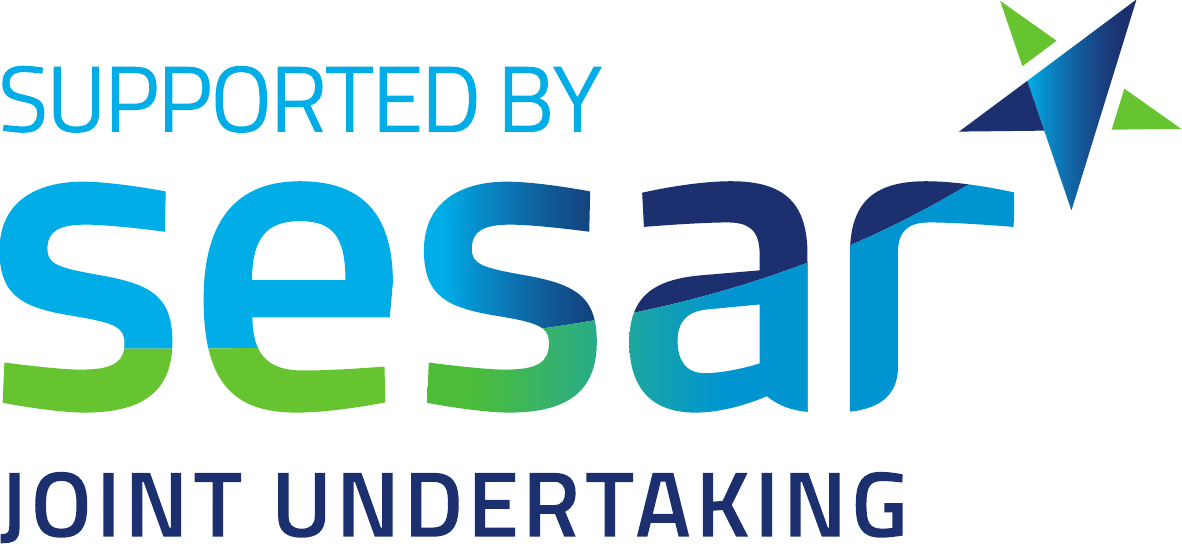}
\end{center}

\section*{Declaration of generative AI and AI-assisted technologies in the manuscript preparation process}
During the preparation of this work the author(s) used 
Claude (Anthropic) and ChatGPT (OpenAI)
in order
to assist with language editing, grammar and spelling correction, and the rephrasing of selected passages to improve clarity and readability.
All scientific ideas, methodology, experimental design, analysis, results, and conclusions were developed by the authors.
The authors reviewed and edited all AI-assisted suggestions and take full responsibility for the content of this publication.

\printcredits
\bibliographystyle{cas-model2-names}
\bibliography{references}


\clearpage

\bio{figs/bashar}
\emph{Bashar Ahmad} is a researcher and lecturer at the Institute of Business Informatics -- Data \& Knowledge Engineering at Johannes Kepler University (JKU) Linz, Austria, where he is a doctoral candidate.
He received his master's degree in computer science and information engineering from National Chiao Tung University in Hsinchu, Taiwan, in 2004, and worked for more than seventeen years in industry as a software engineer and architect before joining JKU Linz.
He is a member of the JKU team in the AWARE project on air traffic management, funded by the SESAR Joint Undertaking under Horizon Europe.
His research interests include knowledge graph management, big data analytics, and applications of artificial intelligence.
\endbio

\bio{figs/christoph}
\emph{Christoph G. Schuetz} is associate professor at the Institute of Business Informatics -- Data \& Knowledge Engineering at Johannes Kepler University (JKU) Linz, Austria. 
He received his doctoral degree from JKU Linz in 2015.
Since 2016, he has been serving as applicant and principal investigator on the JKU side in multiple international collaborative research projects on data management and decision support in air traffic management (ATM) in the Single European Sky ATM Research joint undertaking funded by EU's Horizon frameworks.
His research interests include data engineering and management, business analytics, and applications of artificial intelligence in various domains.
\endbio

\bio{figs/loris}
\textit{Loris Bozzato}
is a researcher at the Department of Theoretical and Applied Sciences (DiSTA) at the University of Insubria in Varese, Italy. 
He received his PhD in 2011 at University of Insubria and, 
from 2011 to 2024, was a researcher at Fondazione Bruno Kessler (FBK) in Trento, Italy.
His research activity focuses on logic for artificial intelligence,
with interest in knowledge representation and reasoning 
and semantic web technologies.
He has co-authored more than 50 publications and is active
in the KR and Semantic Web communities.
\endbio

\end{document}